\documentclass[a4paper,11pt]{article}

\usepackage{jheppub}
\usepackage{here}

\title{
Non-perturbative determination of the $\Lambda$-parameter 
in the pure SU(3) gauge theory 
from the twisted gradient flow coupling 
}

\author[a,b]{Ken-Ichi Ishikawa,}
\author[a]{Issaku Kanamori,}
\author[a]{Yuko Murakami,}
\author[a]{Ayaka Nakamura,}
\author[a,b]{Masanori Okawa}
\author[a]{and Ryoichiro Ueno}

\affiliation[a]{Graduate School of Science, Hiroshima University,\\ Higashi-Hiroshima, Hiroshima 739-8526, Japan}
\affiliation[b]{Core of Research for the Energetic Universe, Hiroshima University,\\ Higashi-Hiroshima, Hiroshima 739-8526, Japan}

\emailAdd{ishikawa@theo.phys.sci.hiroshima-u.ac.jp}
\emailAdd{kanamori@hiroshima-u.ac.jp}
\emailAdd{d152338@hiroshima-u.ac.jp}
\emailAdd{okawa@sci.hiroshima-u.ac.jp}
\emailAdd{ryoichiro-ueno@hiroshima-u.ac.jp}

\preprint{%
{\flushright 
HUPD-1702\\
}}

\newcommand{\TGF}{\mathrm{TGF}}
\newcommand{\SF}{\mathrm{SF}}
\newcommand{\MSbar}{\overline{\mathrm{MS}}}
\newcommand{\Lmax}{L_{\mathrm{max}}}
\newcommand{\LambdaTGF}{\Lambda_{\TGF}}
\newcommand{\LambdaSF}{\Lambda_{\SF}}
\newcommand{\LambdaMSbar}{\Lambda_{\MSbar}}

\abstract{
We evaluate the $\Lambda$-parameter in the $\overline{\mathrm{MS}}$ scheme for the pure SU(3) 
gauge theory with the twisted gradient flow (TGF) method.
A running coupling constant $g_{\TGF}^2(1/L)$ is defined in a finite volume box 
with size of $L^4$ with the twisted boundary condition.
This defines the TGF scheme.
Using the step scaling method for the TGF coupling with lattice simulations, 
we can evaluate the $\Lambda$-parameter non-perturbatively in the TGF scheme.
In this paper we determine the dimensionless ratios, $\Lambda_{\mathrm{TGF}}/\sqrt{\sigma}$ and $r_{0}\Lambda_{\mathrm{TGF}}$
together with the $\Lambda$-parameter ratio $\Lambda_{\mathrm{SF}}/\Lambda_{\mathrm{TGF}}$ on the lattices numerically.
Combined with the known ratio $\Lambda_{\overline{\mathrm{MS}}}/\Lambda_{\mathrm{SF}}$, 
we obtain
$\LambdaMSbar/\sqrt{\sigma} = 0.517(10)(^{+8}_{-7})$ and
$r_{0}\LambdaMSbar=0.593(12)(^{+12}_{-9})$, where the first error is statistical one 
and the second is our estimate of systematic uncertainty.
}

\begin{document} 
\maketitle
\flushbottom


\section{Introduction}
\label{sec:intro}

The $\Lambda$-parameter is a fundamental quantity in asymptotically free gauge theories
and plays the role to set the scale of the theory. 
$\Lambda$ characterizes the low energy non-perturbative physics and
its determination is one of the most important tasks in lattice gauge theory. 
In the pure Yang-Mills theory, $\Lambda$ is the only free parameter of the theory and 
is determined from the coupling constant.
Its value depends on the renormalization scheme. 
In the $\MSbar$ scheme, for example, it is defined by
\begin{equation}
	\LambdaMSbar=\mu(b_{0}g^{2}_{\MSbar}(\mu))^{-\frac{b_{1}}{2b_{0}^{2}}}
	\exp\left[-\frac{1}{2b_{0}g^{2}_{\MSbar}(\mu)}\right]
	\exp\left[-\int_{0}^{g_{\MSbar}(\mu)}\mathrm{d}\xi\left(\frac{1}{\beta(\xi)}+\frac{1}{b_{0}\xi^{3}}-\frac{b_{1}}{b_{0}^{2}\xi}\right)\right],
	\label{def:LambdaMS}
\end{equation}
where $g^{2}_{\MSbar}(\mu)$ is the $\MSbar$ coupling renormalized at the renormalization scale $\mu$, 
and $\beta(\xi)$ is the beta function in the $\MSbar$ scheme. 
 $b_{0}$ and $b_{1}$ are the first two coefficients of the perturbative beta function,
 $b_{0}=\frac{11}{3}\frac{N_{\mathrm{C}}}{16\pi^{2}}$ and $b_{1}=\frac{34}{3}\left(\frac{N_{\mathrm{C}}}{16\pi^{2}}\right)^{2}$, 
for the pure SU($N_{\mathrm{C}}$) gauge theory. 
Since the $\MSbar$ scheme is only defined perturbatively, 
the non-perturbative estimate of $\LambdaMSbar$  thoroughly within the $\MSbar$ scheme is impossible.
Therefore we usually convert a $\Lambda$-parameter determined with a non-perturbative scheme to 
$\LambdaMSbar$ through the perturbative relation.

On the lattice, the $\Lambda$-parameter can be defined by
\begin{align}
	\Lambda_{\mathrm{Lat}}=\frac{1}{a}(b_{0}g^{2}_{\mathrm{Lat}}(1/a))^{-\frac{b_{1}}{2b_{0}^{2}}}&
	\exp\left[-\frac{1}{2b_{0}g^{2}_{\mathrm{Lat}}(1/a)}\right]
	\nonumber \\
	\times&\exp\left[-\int_{0}^{g_{\mathrm{Lat}}(1/a)}\mathrm{d}\xi\left(\frac{1}{\beta(\xi)}+\frac{1}{b_{0}\xi^{3}}-\frac{b_{1}}{b_{0}^{2}\xi}\right)\right],
	\label{def:Lambda}
\end{align}
with the lattice spacing $a$. 
The bare coupling $g_{0}$ can be related to the lattice spacing $a$ non-perturbatively 
and be used as $g_{\mathrm{Lat}}(1/a)=g_{0}$ in eq.~(\ref{def:Lambda}). 
This defines a lattice scheme. 
It is, however, well known that the scaling is largely violated for the range of $g^{2}_{0}$ accessible 
with the presently available computational power. 
In the early stage of the lattice studies, 
it was common to use an improved coupling such as 
$g^{2}_{\mathrm{E}}=\frac{8N_{\mathrm{C}}}{N_{\mathrm{C}}^{2}-1}(1-u_{\mathrm{p}})$ or 
$g^{2}_{\mathrm{A}}=g^{2}_{0}/u_{\mathrm{p}}$ 
with $u_{\mathrm{p}}$ the observed plaquette value~\cite{Allton/Teper/Trivini:StringTension,Parisi:coupling}. 
They exhibit a better scaling property, nonetheless there are only intuitive arguments 
of ``tad-pole improvement'' to explain why they work.

Great progresses for evaluating non-perturbatively running coupling constants have been made 
with the discovery of the step scaling method~\cite{Luscher/Weisz/Wolff:SSF}, 
where the renormalization scale is introduced by the physical box-size of the target system. 
In this method, one can calculate the running coupling in a wide range of the scale covering both the hadronic scale,
 where we make the non-perturbative calculation of physical quantities with lattice techniques,
 and the high energy scale, where we can estimate the $\Lambda$-parameter neglecting higher order corrections. 
The most successful non-perturbative scheme for the running coupling constant in QCD is 
the Schr\"{o}dinger functional (SF) scheme~\cite{SFREFS,SFREFS2,SFREFS3,SFREFS4,SFREFS5,SFREFS6}, 
in which a specific Dirichlet boundary condition is imposed on the temporal direction of the box.
The advantages of the SF scheme are that it is regularization independent and can be defined non-perturbatively.
In addition, the calculation of the $\Lambda$-parameter ratio $\LambdaMSbar/\LambdaSF$ 
has been done in ref.~\cite{Sint/Sommer:LambdaSF} perturbatively.
The disadvantage of the SF scheme, on the other hand, is that it becomes difficult to calculate 
the coupling at larger physical box sizes (i.e. low energy renormalization scale)
due to the appearance of exceptional configurations and the noisy behavior 
which result the large statistical error~\cite{SFREFS3}.

Several other schemes are also available to define the running coupling with the step scaling method. 
The gradient flow scheme is one of the applications of the gradient flow method,
 in which the gauge field is smeared with the so-called flow equation and the smeared gauge field 
has a nice perturbative property on the renormalizability~\cite{Narayanan:2006rf,Luscher:WilsonFlow,Luscher/Weisz:GF}.
In ref.~\cite{Fodor/Holland/Kuti/Nogradi:pGF}, a renormalized coupling via the gradient flow 
 in a finite size box with the periodic boundary condition has been introduced. 
However, $\LambdaMSbar$ cannot be extracted from the coupling, 
since the coupling has a non-analytic expansion in $\alpha_{\MSbar}$ due to the zero-mode of the gauge 
field in the periodic boundary condition. 
To avoid the zero-mode problem, the twisted boundary condition has been introduced by Ramos~\cite{Ramos:TGF}. 
The renormalized coupling defined in a finite box with the twisted boundary condition (the TGF scheme) has 
the normal one-loop relation to the $\MSbar$ scheme and is regularization independent. 
The running can be traced via the step scaling method on the lattice. 
The TGF running coupling for the pure SU(2) Yang-Mills theory has been evaluated using 
the step scaling method~\cite{Ramos:TGF} 
and extended to two-color many flavor dynamical simulations~\cite{Lin:2015zpa}. 
The gradient flow coupling with the Schr\"{o}dinger functional boundary condition is another scheme 
avoiding the zero-mode problem and has been investigated 
in refs.~\cite{Fritzsch/Ramos:GF,Brida:2015gqj,Leino:2015bfg} for the SU(3) gauge theories.

We extend Ramos's work~\cite{Ramos:TGF} to the pure SU(3) Yang-Mills theory. 
In addition to this, we extract the $\Lambda$-parameter in the TGF scheme and convert it to the $\MSbar$ scheme. 
The ratio $\LambdaMSbar/\LambdaTGF$, which is usually evaluated using the perturbation theory,
 is not yet available at this time (but there is an ongoing study~\cite{bribian}). 
Since we already know $\LambdaMSbar/\LambdaSF$~\cite{Sint/Sommer:LambdaSF}, 
 actually what we have to estimate is the ratio $\LambdaSF/\LambdaTGF$. 
Therefore we estimate $\LambdaSF/\LambdaTGF$ for the pure SU(3) gauge theory numerically 
with lattice simulations in this study. 
It should be noted that the analysis made in this paper is applicable to the gauge theories 
with dynamical fermions provided that the fermion representations and contents are compatible 
with the twisted boundary condition. 
This study is the first attempt to apply the TGF method for evaluating 
the $\Lambda$-parameter in the SU(3) gauge theories from the beginning to the end.

In this study we estimate $\LambdaMSbar$ in terms of physical observables via the TGF method. 
Our strategy is summarized as follows:
\begin{equation}
  \frac{\LambdaMSbar}{A_{\mathrm{phys}}}=
   \frac{\LambdaMSbar}{\LambdaSF}  \cdot
      \frac{\LambdaSF}{\LambdaTGF} \cdot
   \frac{\Lmax \LambdaTGF}{\Lmax A_{\mathrm{phys}}}.
  \label{strategy}
\end{equation}
Here $A_{\mathrm{phys}}$ is a physical observable with mass dimension and 
 $\Lmax $ is an intermediate scale which connects the non-perturbative energy scale and the perturbative energy scale.
In this paper, we consider the string tension $\sqrt{\sigma}$ and the Sommer scale $1/r_{0}$ 
as the physical observable $A_{\mathrm{phys}}$. 
(Another reference scale can be considered, for example $w_{0}$~\cite{refscale}.) 
We will numerically calculate $\Lmax \LambdaTGF$, $\Lmax A_{\mathrm{phys}}$, and $\LambdaSF/\LambdaTGF$. 
$\Lmax\LambdaTGF$ is calculated with the step scaling method. 
In order to evaluate $\Lmax A_{\mathrm{phys}}$,
we employ data available 
from refs.~\cite{Allton/Teper/Trivini:StringTension,Antonio/Okawa:StringTension} 
 and ref.~\cite{Necco:Dthesis}
 for $a\sqrt{\sigma}$ and $a/r_{0}$, respectively.
We finally estimate $\LambdaMSbar/A_{\mathrm{phys}}$ using eq.~(\ref{strategy}). 
We show that our estimates for $\LambdaMSbar/A_{\mathrm{phys}}$ are compatible with 
the values previously obtained with other methods. 
This demonstrates the validity of our non-perturbative analysis.

This paper is organized as follows. 
In the next section, we introduce the TGF method and explain how to calculate the TGF coupling briefly. 
Our strategy eq.~(\ref{strategy}) and the details of lattice simulations are explained in section~\ref{sec:strategy}. 
$\Lmax \LambdaTGF$ and 
$\Lmax A_{\mathrm{phys}}$ are presented in sections~\ref{sec:LmbdaTGF} and \ref{sec:STSS}, respectively.
$\LambdaSF/\LambdaTGF$ and $\LambdaMSbar/A_{\mathrm{phys}}$ are extracted in section~\ref{sec:ratio}. 
Finally we summarize this paper in the last section~\ref{sec:summary}. 
Our preliminary result has been presented at the Lattice conference~\cite{ours}.


\section{Twisted gradient flow coupling}
\label{sec:tgf}

We use the Wilson gauge action on a $(L/a)^{4}$ lattice with twisted boundary condition:
\begin{equation}
  S_{\mathrm{W}}[U] = 
    \frac{\beta}{2N_{\mathrm{C}}}
    \sum_{{n,\mu,\nu}\atop{(\mu\neq\nu)}}
        Z_{\mu\nu}(n)\mathrm{Tr}\left[U_{\mu}(n)U_{\nu}(n+\hat{\mu})U_{\mu}^{\dagger}(n+\hat{\nu})U_{\nu}^{\dagger}(n)\right].
  \label{WilsonAction}
\end{equation}
Here $U_{\mu}(n)$ is the SU($N_{\mathrm{C}}$) link variable with \emph{periodic} boundary condition. 
We represent the twisted boundary condition by using the twist phase $Z_{\mu\nu}(n)$.
In this work, we follow ref.~\cite{Ramos:TGF} and put the twisted boundary condition in the $x$-$y$ plane.
The twist phase is defined as
\begin{align}
  Z_{\mu\nu}(n)=Z_{\nu\mu}^{\ast}(n)=
  \begin{cases}
    \exp\left[-\frac{2\pi i}{N_{\mathrm{C}}}\right] & \mu=1,\nu=2,\ \text{and}\ n_{1}=n_{2}=0, \\
     1                                              & \text{otherwise},
  \end{cases}
\end{align}
in the case.
The derivation of the action with the periodic variables~(\ref{WilsonAction}) is given in appendix~\ref{app:A}.

We first introduce link variables $V_{\mu}(n,t)$ evolved with the gradient flow equation;
\begin{equation}
  \frac{\mathrm{d}V_{\mu}(n,t)}{\mathrm{d}t} = 
          -\frac{2N_{\mathrm{C}}}{\beta}\left\{\partial_{n,\mu}S_{\mathrm{W}}[V]\right\}V_{\mu}(n,t),
                 \quad  \left.V_{\mu}(n,t)\right|_{t=0}=U_{\mu}(n),
  \label{FlowEq'}
\end{equation}
where $t$, a fictitious time or so called flow time, is introduced.
$\partial_{n,\mu}$ is the $\mathfrak{su}(N_{\mathrm{C}})$-valued differential operator 
with respect to $V_{\mu}(n,t)$.

The twisted gradient flow (TGF) coupling $g^{2}_{\TGF}(1/L)$ is defined as
\begin{equation}
  g_{\TGF}^{2}(1/L) = \left.\mathcal{N}^{-1}_{\mathrm{T}}(c,a/L)t^{2}\langle E(t)\rangle\right|_{t=c^{2}L^{2}/8},
  \label{TGFcoupling}
\end{equation}
where $E(t)$ is a energy density made of $V_{\mu}(n,t)$. The explicit form of $E(t)$ will be given later.
The vacuum expectation value $\langle E(t)\rangle$ is a renormalized quantity 
at the scale $1/\sqrt{8t}$ at any $t>0$~\cite{Luscher/Weisz:GF}.
In a finite volume system we can use the volume size $L$ as the scale of the renormalization so 
we have set
\begin{equation}
	\frac{1}{\sqrt{8t}}=\frac{1}{cL}
	\label{FlowScale}
\end{equation}
in eq.~(\ref{TGFcoupling}).
The factor $c$ is, in principle, a free parameter: a different choice of $c$ gives a different renormalization scheme. 
Throughout this work we choose $c=0.3$ for a reason we will state later.
The normalization factor $\mathcal{N}^{-1}_{\mathrm{T}}(c,a/L)$ depends on the definition of the energy density on the lattice.

In this work, we employ the following definition for the energy density $E(t)$;
\begin{equation}
  E(t)=-\frac{1}{64N_{\mathrm{C}}(L/a)^{4}}\sum_{n,\mu\neq\nu}\mathrm{Tr}\left[G_{\mu\nu}^{2}(n,t)\right],
  \label{EnergyDensity}
\end{equation}
with
\begin{align}
 G_{\mu\nu}(n,t)&
   = Z_{\mu\nu}(n)V_{\mu}(n,t)V_{\nu}(n+\hat{\mu},t)V_{\mu}^{\dagger}(n+\hat{\nu},t)V_{\nu}^{\dagger}(n,t)
  \notag
  \\
  &+Z_{\mu\nu}(n-\hat{\mu})V_{\nu}(n,t)V_{\mu}^{\dagger}(n-\hat{\mu}+\hat{\nu},t)V_{\nu}^{\dagger}(n-\hat{\mu},t)V_{\mu}(n-\hat{\mu},t)
  \notag
  \\
  &+Z_{\mu\nu}(n-\hat{\mu}-\hat{\nu})V_{\mu}^{\dagger}(n-\hat{\mu},t)V_{\nu}^{\dagger}(n-\hat{\mu}-\hat{\nu},t)V_{\mu}(n-\hat{\mu}-\hat{\nu},t)V_{\nu}(n-\hat{\nu},t)
  \notag
  \\
  &+Z_{\mu\nu}(n-\hat{\nu})V_{\nu}^{\dagger}(n-\hat{\nu},t)V_{\mu}(n-\hat{\nu},t)V_{\nu}(n+\hat{\mu}-\hat{\nu},t)V_{\mu}^{\dagger}(n,t)-\{\mathrm{h.c.}\}.
  \label{FieldStrength}
\end{align}
With this definition, the normalization factor $\mathcal{N}^{-1}_{\mathrm{T}}(c,a/L)$, 
which is defined so as to match $g_{\TGF}^{2}(1/L)$ with the bare coupling $g_{0}^{2}$ at the tree level 
of the perturbation theory, is
\begin{equation}
  \mathcal{N}^{-1}_{\mathrm{T}}(c,a/L) = 
   \frac{c^{4}}{128}\sum_{P}^{\prime}
     \exp\left[{-\frac{c^{2}L^{2}}{4}\hat{P}^{2}}\right]\frac{\tilde{P}^{2}C^{2}-(\tilde{P}_{\mu}C_{\mu})^{2}}{\hat{P}^{2}},
  \label{norm'factor}
\end{equation}
where 
\begin{equation}
    \hat{P}_{\mu}=\frac{2}{a}\sin\left[a\frac{P_{\mu}}{2}\right],\quad
  \tilde{P}_{\mu}=\frac{1}{a}\sin[aP_{\mu}],\quad 
          C_{\mu}=\cos\left[a\frac{P_{\mu}}{2}\right].
	\label{lattice-momentum}
\end{equation}
The summation over $P_{\mu}$ runs
\begin{equation}
	P_{1,2}=\frac{2\pi m_{1,2}}{N_{\mathrm{C}}L},\quad 0\leq m_{1,2}\leq\frac{N_{\mathrm{C}}L}{a}-1,
	\label{sum'1-2}
\end{equation}
for $\mu=1,2$ and
\begin{equation}
	P_{3,4}=\frac{2\pi m_{3,4}}{L},\quad 0\leq m_{3,4}\leq \frac{L}{a}-1,
	\label{sum'3-4}
\end{equation}
for $\mu=3,4$.
The prime ($\prime$) symbol on the summation indicates 
the exclusion of the zero momentum contribution $(P_1,P_2,P_3,P_4)=(0,0,0,0)$ from the sum.

We employ $c=0.3$ throughout this work. 
In general, a smaller value of $c$ gives smaller statistical error. 
It causes, however, a larger lattice artifact. 
According to the previous works~\cite{Ramos:TGF,Fritzsch/Ramos:GF}, $c=0.3$ gives a good compromise 
between these two effects. This is the reason for our choice $c=0.3$.


\section{Overview of strategy and simulation details}
\label{sec:strategy}

Here we explain the strategy for evaluating eq.(\ref{strategy}).
We take the following steps.
\begin{enumerate}
  \item
    We evaluate the discrete beta function $B_{s}(u)$ as a function of $u=g_{\TGF}^{2}(1/L)$. 
    It is defined as
    \begin{equation}
      B_{s}\left(g_{\TGF}^{2}(1/L)\right)=\dfrac{g_{\TGF}^{2}(s/L)-g_{\TGF}^{2}(1/L)}{\log[s^{2}]},
      \label{eq:DiscBetaF}
    \end{equation}
    where $s$ is the scaling parameter. 
    We extract this discrete beta function by taking the continuum limit of 
    lattice discrete beta functions evaluated on several lattices.
    The details of the fitting and the analysis for the continuum limit will be explained in the next section.
  \item
    We estimate $\Lmax  \LambdaTGF$ using the discrete beta function evaluated in the previous step. 
    By fixing the scale $1/\Lmax $ implicitly through the value of the coupling $u^{\ast}=g_{\TGF}^{2}(1/\Lmax )$, 
    $\Lmax\LambdaTGF$ can be evaluated with 
    \begin{align}
       c\Lmax \LambdaTGF& = \left(b_{0}u^{\ast}\right)^{-\frac{b_{1}}{2b_{0}^{2}}}
                             \exp\left[-\frac{1}{2b_{0}u^{\ast}}\right]
       \notag\\
                        &\qquad 
                           \times\exp
                            \left[
                              -\int_{0}^{\sqrt{u^{\ast}}}\mathrm{d}\xi
                                    \left( 
                                            \frac{1}{\beta_{\TGF}(\xi)}+\frac{1}{b_{0}\xi^{3}}
                                            -\frac{b_{1}}{\left(b_{0}\right)^{2}\xi}
                                    \right)
                            \right]
       \notag\\
                        & \simeq s^{n}\left(b_{0}u_{n}\right)^{-\frac{b_{1}}{2b_{0}^{2}}}
                                 \exp\left[-\frac{1}{2b_{0}u_{n}}\right].
       \label{eq:LambdaTGFTwoLoop}
    \end{align}
    Here we explicitly put $c$ on the left-hand side, which is to use the same notation as eq.~(\ref{FlowScale}) for the scale setting.
    The TGF coupling at scale $s^n/\Lmax $ is evaluated with the following recurrence equation (step scaling),
    \begin{equation}
      u_{i}=u_{i-1}+B_{s}(u_{i-1})\log[s^{2}],\qquad u_{0}=u^{\ast}.
      \label{eq:StepScale}
    \end{equation}
    For a sufficiently small value of $u_{n}=g_{\TGF}^{2}(s^{n}/\Lmax )$ 
    we can safely use the two-loop approximation in eq.~(\ref{eq:LambdaTGFTwoLoop}) to extract $\Lmax\LambdaTGF$.
  \item 
    We relate the intermediate scale $1/\Lmax $ to a hadronic scale $A_{\mathrm{phys}}$ in the continuum limit. 
    We employ two hadronic scales for the consistency check; 
      the string tension $\sqrt{\sigma}$ and the Sommer scale $r_{0}$. 
    The lattice data of $\sqrt{\sigma}$ and $r_{0}$ are taken from 
      refs.~\cite{Allton/Teper/Trivini:StringTension,Antonio/Okawa:StringTension} 
        and~\cite{Necco:Dthesis}, respectively. 
    To outline the procedure, let us assume that $A_{\mathrm{phys}}$ has a mass dimension one for simplicity. 
    We interpolate each of $g_{\TGF}^{2}(1/L,\beta)$ and $aA_{\mathrm{phys}}(\beta)$ as a function of bare coupling $\beta$. 
    By keeping the coupling constant $g_{\TGF}^{2}(1/\Lmax,\beta^{\ast})$ fixed to $u^{\ast}$ over several lattices $\Lmax/a^{\ast}$, 
    we obtain the corresponding values of $\beta^{\ast}$ 
    (here to show the connection between $u^{\ast}$ and the lattice spacing (or bare coupling), we use $a^{\ast}$ (or $\beta^{\ast}$) ).
    For each value of $\beta^{\ast}$ (thus $a^{\ast}/\Lmax $) we have a pair of $\Lmax /a^{\ast}$ and $a^{\ast}A_{\mathrm{phys}}(\beta^{\ast})$. 
    We then take the continuum limit of $(\Lmax /a^{\ast})(a^{\ast}A_{\mathrm{phys}})$ as a function of $a^{\ast}/\Lmax$. 
  \item
    To convert $\LambdaTGF$ to the $\Lambda$-parameter in the $\MSbar$ scheme, we need the ratio $\LambdaMSbar/\LambdaTGF$.
    We split the ratio into two pieces: $(\LambdaMSbar/\LambdaSF)(\LambdaSF/\LambdaTGF)$. 
    The value of the former factor is already known to be $\LambdaMSbar/\LambdaSF=0.48811(1)$~\cite{Sint/Sommer:LambdaSF}, 
    but the latter is not known in the literature. 
    We therefore calculate $\LambdaSF/\LambdaTGF$ numerically via the one-loop relation between 
    $g^{2}_{\SF}$ and $g^{2}_{\TGF}$ at the same renormalization scale $1/L$.
    To obtain the one-loop relation, we calculate the couplings with lattice simulations in the weak coupling region.
  \item 
    Finally we combine the all pieces obtained above to have
    \begin{align}
       \frac{\LambdaMSbar}{\sqrt{\sigma}} =
       \frac{\LambdaMSbar}{\LambdaSF}
       \frac{\LambdaSF}{\LambdaTGF}
       \frac{\Lmax \LambdaTGF}{\Lmax \sqrt{\sigma}},
       \quad 
       r_{0}\LambdaMSbar = 
       \frac{\LambdaMSbar}{\LambdaSF}
       \frac{\LambdaSF}{\LambdaTGF}
       \frac{\Lmax \LambdaTGF}{\Lmax /r_{0}}.
    \end{align}
\end{enumerate}


The TGF couplings on the lattice in the steps 1--3 explained above are evaluated on five lattices with $L/a=$12, 16, 18, 24 and 36.
We use the heat-bath method introduced by Fabricius and Haan~\cite{heatbath} to increase the acceptance ratio.
We accumulate configurations as listed in table~\ref{table:number_of_configs_with_ac} in appendix~\ref{app:B}.
Each configuration is separated by 100 sweeps.
The TGF couplings, we computed,  are listed in table~\ref{table:rawdata}, 
 of which error is statistical one 
 and  estimated by taking the autocorrelation into account with the procedure  proposed 
 in ref.~\cite{Wolff:2003sm}\footnote{We observed long autocorrelations for some of the parameter 
      sets so we increased the statistics for them. 
      We leave the identification of the source of this behavior for future study.}. 
We take several values for the bare coupling $\beta=6/g_{0}^{2}$ on each lattice to take the continuum limit.

On the other hand, simulations in the weak coupling region have been done on four lattices $L/a=$8, 10, 12 and 16, 
with three values of the bare coupling $\beta=$ 40, 60 and 80.
We use the same plaquette gauge action with the $O(a)$-improvement boundary correction and 
the SF boundary condition~\cite{SFREFS,SF:two-loop} to calculate $g_{\SF}^{2}$. 
The error of the coupling from these data is estimated with the Jackknife method after binning data into 10 bins.
We execute $O(10^{6})${\bf{--}}$O(10^{7})$ sweeps for each parameter.
The SF coupling is evaluated every sweep and the TGF coupling is evaluated every 100 sweeps.

The error propagation of the statistical error on non-primary observables, 
 such as the discrete beta function in the continuum limit, is estimated by a random re-sampling method.
For the re-sampling, we assume the primary data in table~\ref{table:rawdata} 
satisfies Gaussian distribution with the width of the measured statistical error.

\begin{table}[tb]
	\centering
	\begin{tabular}{|c|cccccc|}
		\hline
		        &             &            & $g_{\TGF}^{2}$      &            &  &
		\\
		$\beta$ &   12        &   16       &   18       &   24       & 36 &  $L/a$
		\\ \hline
		 6.11   & 6.9717(35)  &            &            &            &    &         
		\\
		 6.20   & 5.8715(38)  &            &            &            &    &           
		\\
		 6.29   &             &            & 8.423(28)  &            &    &         
		\\
		 6.30   &             & 7.0234(94) &            &            &    &         
		\\
		 6.38   &             &            & 7.082(14)  &            &    &         
		\\
		 6.40   & 4.5129(29)  & 5.892(12)  & 6.848(14)  &            &    &         
		\\
		 6.50   & 4.1052(26)  & 5.118(13)  & 5.734(14)  &            &    &         
		\\
		 6.60   &             & 4.5733(58) &            & 6.967(31)  &    &         
		\\
		 6.75   & 3.3836(18)  & 3.9874(69) & 4.3092(87) & 5.459(12)  &    &         
		\\
		 6.90   &             &            &            & 4.6212(83) & 7.065(47)  & 
		\\
		 7.00   & 2.9041(14)  & 3.3168(75) & 3.5318(73) & 4.2291(91) & 5.962(44)  & 
		\\
		 7.25   & 2.5489(12)  & 2.8649(40) & 3.0163(57) & 3.4855(78) & 4.467(21)  & 
		\\
		 7.50   & 2.2783(11)  & 2.5122(41) & 2.6370(57) & 2.9838(83) & 3.672(12)  & 
		\\
		 7.75   & 2.06258(94) & 2.2627(32) & 2.3452(47) & 2.6121(73) & 3.126(13)  & 
		\\
		 8.00   & 1.88717(88) & 2.0469(36) & 2.1238(42) & 2.3304(54) & 2.715(13)  & 
		\\
		 8.25   & 1.74054(80) & 1.8722(30) & 1.9380(35) & 2.1109(47) & 2.4035(77) & 
		\\
		 8.50   & 1.61638(71) & 1.7312(26) & 1.7868(34) & 1.9264(40) & 2.1657(66) & 
		\\
		 9.00   & 1.41544(65) & 1.4971(24) & 1.5438(39) & 1.6460(43) & 1.8190(70) & 
		\\
		 9.50   & 1.26024(62) & 1.3242(22) & 1.3576(30) & 1.4391(46) & 1.5748(54) & 
		\\
		10.00   & 1.13594(49) & 1.1879(16) & 1.2134(17) & 1.2804(33) & 1.3759(51) & 
		\\ \hline
	\end{tabular}
	\caption{TGF coupling on each $L/a$ and $\beta$.}
	\label{table:rawdata}
\end{table}


\section{TGF running coupling constant and $\Lmax \LambdaTGF$}
\label{sec:LmbdaTGF}

\newcommand{\figscale}{0.4}

\begin{figure}[t]
  \begin{tabular}{c}
    \begin{minipage}{1.0\hsize}
      \centering
      \includegraphics[scale=\figscale,angle=-90,clip]{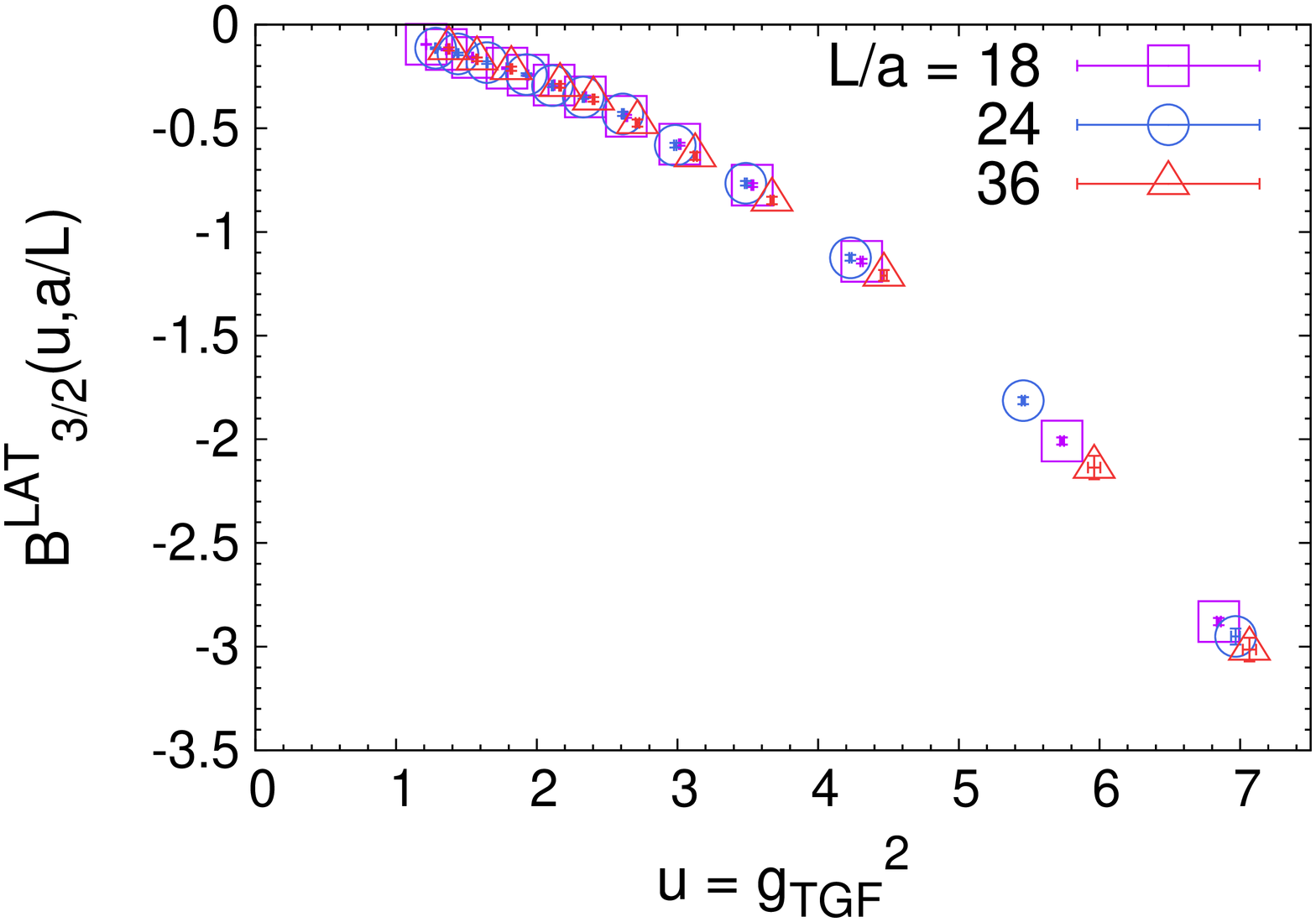}
      \caption{The discrete beta function on each lattice size.}
      \label{fig:B1}
    \end{minipage}
  \\ \ \\
    \begin{minipage}{1.0\hsize}
      \centering
      \includegraphics[scale=\figscale,angle=-90,clip]{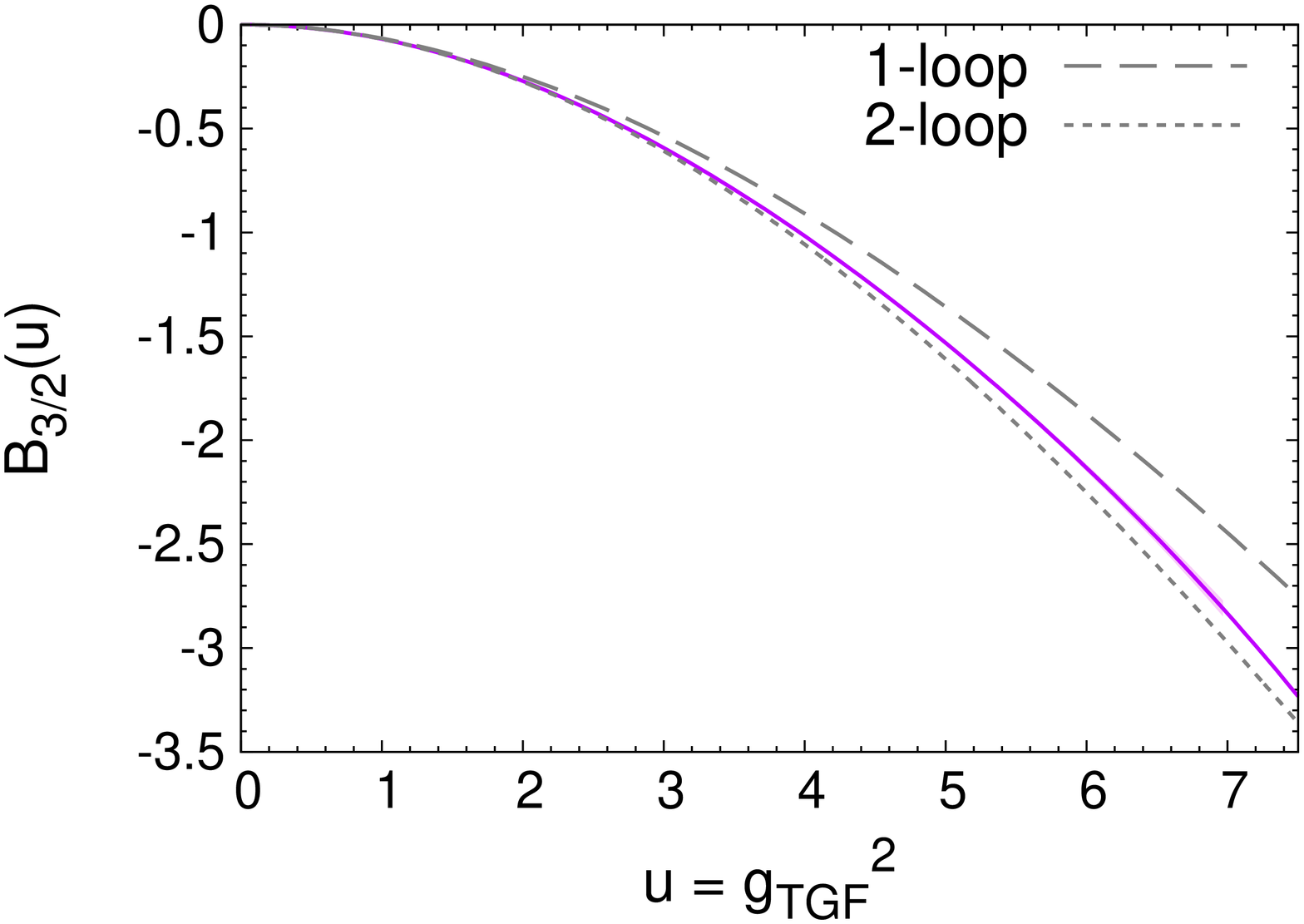}
      \caption{The discrete beta function in the continuum limit (solid purple line) 
               together with the one- and two-loop analytic results.  
               The statistical error band (light purple shade) for the result is 
               underlaid behind the solid purple line, but is 
              almost same as the width of the line.}
      \label{fig:B2}
    \end{minipage}
  \end{tabular}
\end{figure}

To extract the discrete beta function eq.~(\ref{eq:DiscBetaF}), we take the continuum limit of the lattice discrete beta function defined by
\begin{align}
  B_s^{\mathrm{LAT}}(g_{\TGF}^{2}(1/L,\beta)) =\dfrac{g_{\TGF}^{2}(s/L,\beta)-g_{\TGF}^{2}(1/L,\beta)}{\log[s^{2}]}.
\label{eq:DiscBetaLAT}
\end{align}
We use $s=3/2$ as the scaling parameter. 
To take the continuum limit of eq.~(\ref{eq:DiscBetaLAT}), 
 the value of $g_{\TGF}^{2}(1/L,\beta)$ is kept fixed at $g_{\TGF}^{2}(1/L,\beta)=u$ 
 as the renormalization condition irrespective of $\beta$.
This implies that the physical length $L$ is fixed.
The lattice discrete beta function is evaluated using eq.~(\ref{eq:DiscBetaLAT}) 
by substituting the data of table~\ref{table:rawdata}.
We fit the lattice discrete beta function with
\begin{align}
  B_{3/2}^{\mathrm{LAT}}(u,a/L) = & 
  \left[\sigma_{0}+d_{1}\left(\dfrac{a}{L}\right)^{2}\right]u^{2}+\left[\sigma_{1}+d_{2}\left(\dfrac{a}{L}\right)^{2}\right]u^{3}
 \notag \\
    &+\left[c_{3}+d_{3}\left(\dfrac{a}{L}\right)^{2}\right]u^{4}+\left[c_{4}+d_{4}\left(\dfrac{a}{L}\right)^{2}\right]u^{5},
\label{eq:FitDiscBeta}
\end{align}
as a function of $a/L$ and $u$. 
Here $\sigma_{0}=-b_{0}$ and $\sigma_{1}=\sigma_{0}^{2}\log[s^{2}]-b_{1}$ 
are fixed to the analytical values from the two-loop perturbation. 
We fit all data $(B_{3/2}^{\mathrm{LAT}}(u,a/L),u, a/L)$ simultaneously 
by taking the correlation among $u$'s and $B_{3/2}^{\mathrm{LAT}}(u,a/L)$'s into account~\cite{AWAYA:1982}.
The continuum limit is obtained by dropping the $d_{j}$ terms.
The fit result is
\begin{align}
  &B_{3/2}^{\mathrm{LAT}}(u,a/L)=\left[\sigma_{0}+0.19(52)(a/L)^{2}\right]u^{2}+\left[\sigma_{1}+1.19(47)(a/L)^{2}\right]u^{3}
  \notag \\
  &\quad
    +\left[0.000624(44)-0.39(14)(a/L)^{2}\right]u^{4}+\left[-0.0000515(76)+0.027(12)(a/L)^{2} \right]u^{5},
  \label{eq:disc-beta}
\end{align}
with $\chi^{2}/\mathrm{DoF}=0.94(45)$.
Figures~\ref{fig:B1} and~\ref{fig:B2} show $B_{s}^{\mathrm{LAT}}$ and $B_{s}$ respectively.
Plotted are also analytic one- and two-loop lines for comparison.

We evaluate $\Lmax \LambdaTGF$ according to the step 2 in section~\ref{sec:strategy}.
Eleven values from $u^{\ast}=6.0,6.1,\dots,7.0$ are taken for the intermediate scale $\Lmax$.
The fluctuations coming from the different choice of $\Lmax$ will be used to estimate the systematic errors 
of the final results of $r_{0}\LambdaMSbar$ and $\LambdaMSbar/\sqrt{\sigma}$ in section~\ref{sec:STSS}.
The number of steps $n$ to evolve eq.~(\ref{eq:StepScale}) is $n=200$, 
 where $u_{n=200}$ is sufficiently small to utilize eq.~(\ref{eq:LambdaTGFTwoLoop}). 
The values of $c\Lmax  \LambdaTGF$ for each $u^{\ast}$ are tabulated in table~\ref{table:LambdaTGF}.

\begin{table}[t]
	\centering
	\begin{tabular}{|cc|}
		\hline
		$u^{\ast}=g_{\TGF}^{2}(1/\Lmax )$ & $c\Lmax \LambdaTGF$
		\\ \hline
		6.0                                         & 0.570(10)
		\\
		6.1                                         & 0.579(10)
		\\
		6.2                                         & 0.588(11)
		\\
		6.3                                         & 0.597(11)
		\\
		6.4                                         & 0.605(11)
		\\
		6.5                                         & 0.613(11)
		\\
		6.6                                         & 0.621(12)
		\\
		6.7                                         & 0.629(12)
		\\
		6.8                                         & 0.636(12)
		\\
		6.9                                         & 0.643(12)
		\\
		7.0                                         & 0.650(13)
		\\ \hline
	\end{tabular}
	\caption{$c\Lmax \LambdaTGF$ for each $u^{\ast}$.}
	\label{table:LambdaTGF}
\end{table}


\section{Physical scale in terms of $\Lmax $}
\label{sec:STSS}

As described in section~\ref{sec:strategy}, the hadronic scales, 
the string tension $\sqrt{\sigma}$ and the Sommer scale $r_{0}$,
have to be determined in terms of $\Lmax$.
$a\sqrt{\sigma}$ and $r_{0}/a$ with the plaquette gauge action 
in large physical volumes
have been determined 
 at $\beta\in[5.65,6.515]$ in refs.~\cite{Allton/Teper/Trivini:StringTension,Antonio/Okawa:StringTension} 
and $\beta\in[5.70,6.692]$ in ref.~\cite{Necco:Dthesis}, respectively. 
To relate the intermediate scale $\Lmax /a$ and the physical scales 
$a A_{\mathrm{phys}}$ ($= a\sqrt{\sigma}$ or $a/r_{0}$) 
at the same lattice cut-off ``$a$'', 
we need the bare coupling constant $g_{0}^{2}$ dependence (or $\beta$ dependence) 
of $g^{2}_{\TGF}(1/\Lmax,\beta)$ and $a A_{\mathrm{phys}}(\beta)$. 

If the values of $aA_{\mathrm{phys}}(\beta^{\ast})$ 
at a fixed value $g^{2}_{\TGF}(a^{\ast}/\Lmax ,\beta^{\ast})=u^{\ast}$
 on several lattice sizes are obtained, we can take the continuum limit 
for $\Lmax  A_{\mathrm{phys}}$ as follows:
\begin{align}
  \Lmax  A_{\mathrm{phys}} = 
    \left.\lim_{a^{\ast}/\Lmax \to0}
         \left[
           \left(\frac{\Lmax}{a^{\ast}}\right)
           \left(a^{\ast}A_{\mathrm{phys}}\right)
         \right]\right|_{\mathrm{fixed\ }g^{2}_{\TGF}=u^{\ast}}.
  \label{eq:cont_LmaxAphys}
\end{align}
To take the continuum limit of the hadronic scale $aA_{\mathrm{phys}}$ reliably,
  $g_{\TGF}^{2}(1/\Lmax ,\beta)$ should be precisely evaluated in the scaling 
  region of $aA_{\mathrm{phys}}$ on several lattice sizes $\Lmax/a$ with sufficiently large $u^{\ast}$.
This condition is satisfied with our data at $\Lmax /a=12,16$ and $18$, 
where the large enough TGF couplings $g^{2}_{\TGF}(1/\Lmax,\beta)=u^{\ast}$ and $aA_{\mathrm{phys}}$ 
in the scaling region are available in the ranges 
    $\beta\in[6.11,6.515]$ for $a\sqrt{\sigma}$
and $\beta\in[6.11,6.92]$  for $a/r_{0}$, respectively.
Therefore we can take any renormalization condition $u^{\ast}$ in this region and 
we employ several different values $u^{\ast}=6.0,6.1\dots,7.0$ to see 
the consistency as stated in the previous section.

Let us start with interpolation of $g^{2}_{\TGF}(1/\Lmax,\beta)$, $a\sqrt{\sigma}$ and $a/r_{0}$ as 
functions of $\beta$ separately in the following.
Then we combine the interpolated results to take the continuum limit 
using eq.~(\ref{eq:cont_LmaxAphys}).

\begin{figure}[t]
  \begin{tabular}{c}
    \begin{minipage}{1.0\hsize}
      \centering
      \includegraphics[scale=\figscale,angle=-90,clip]{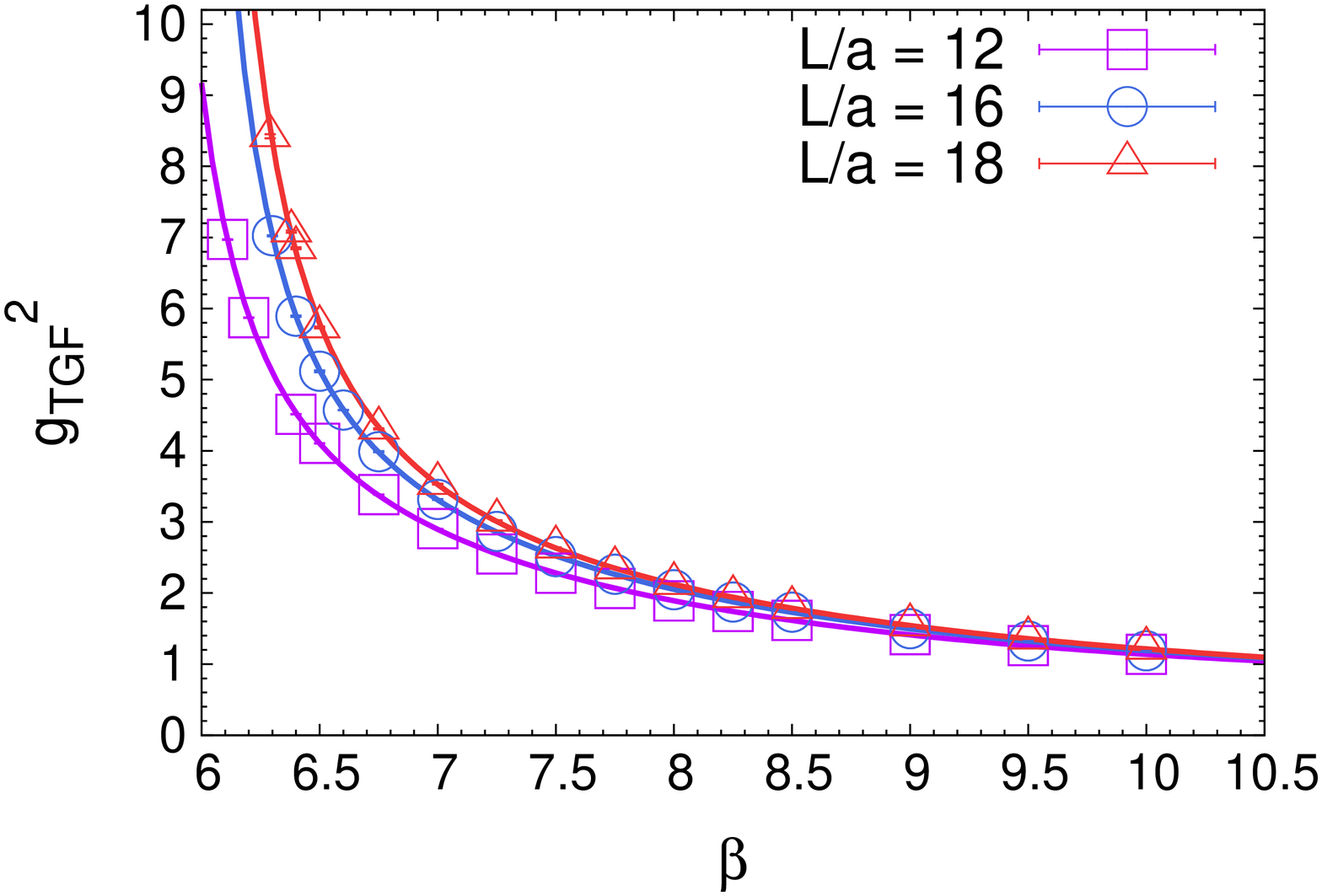}
      \caption{$g_{\TGF}^2(1/L,\beta)$ vs $\beta$ at each lattice size. 
               The solid lines show the fit results with eq.~(\ref{eq:interplating_gTGF}).}
      \label{figure:gTGFfit}
    \end{minipage}
  \\ \ \\
    \begin{minipage}{1.0\hsize}
      \centering
      \includegraphics[scale=\figscale,angle=-90,clip]{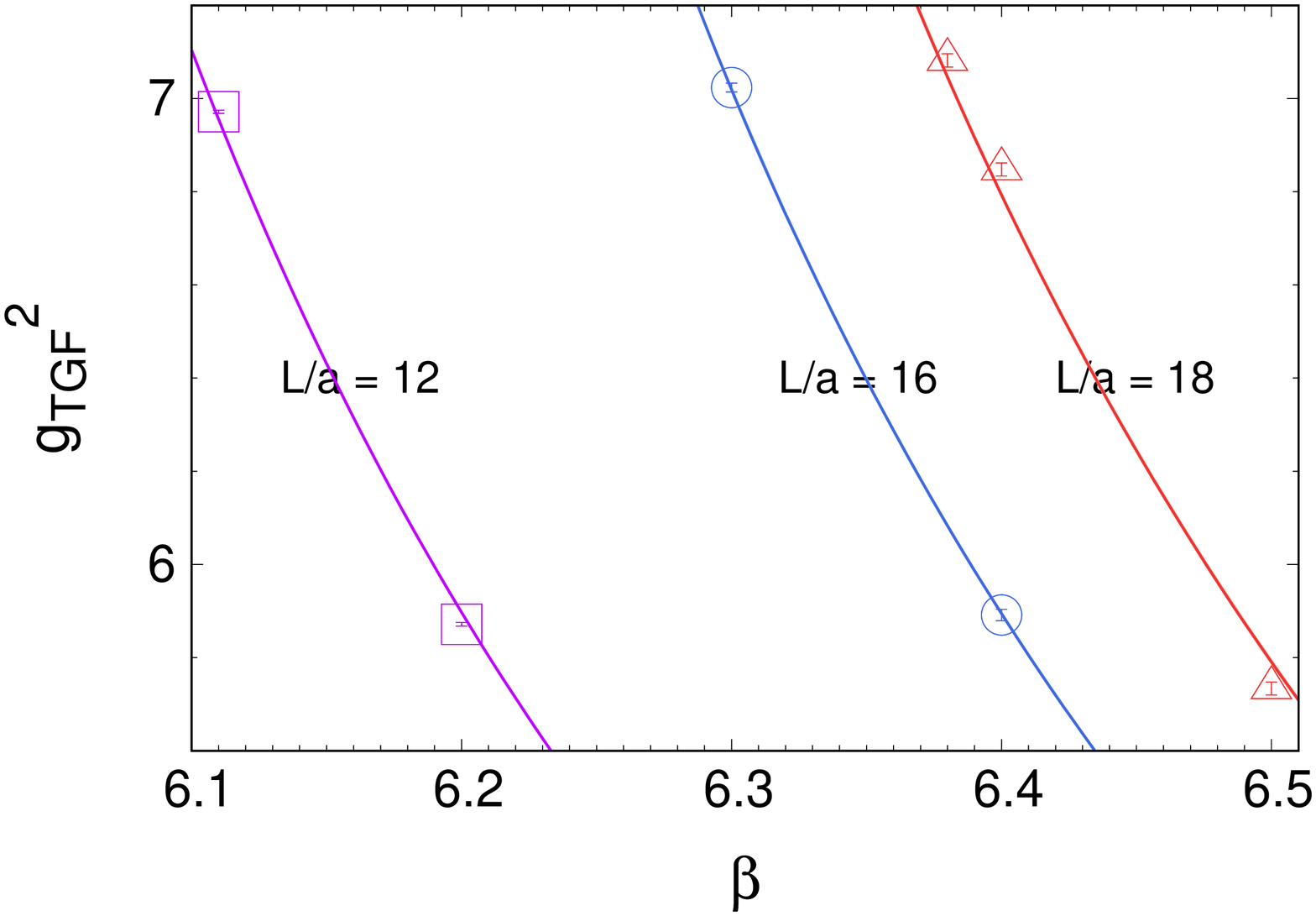}
      \caption{Magnification of figure~\ref{figure:gTGFfit}.}
      \label{figure:gTGFfit_en}
    \end{minipage}
  \end{tabular}
\end{figure}

To interpolate $g^{2}_{\TGF}(1/\Lmax ,\beta)$, we fit the data at $\Lmax/a=12$, $16$ and $18$ 
in table~\ref{table:rawdata} with the following interpolating function;
\begin{align}
  g^{2}_{\TGF}(1/\Lmax ,\beta)=g_{0}^{2}\dfrac{1}{1+\sum_{k=1}^{5}c_{i}g_{0}^{2i}}.
 \label{eq:interplating_gTGF}
\end{align}
We use all the data in $\beta\in[6.1,10.0]$ to stabilize the interpolation, 
while we need the interpolating formula only in the scaling region corresponding to $u^{\ast}$ we chose.
Figure~\ref{figure:gTGFfit} shows the fit result, and table~\ref{table:gTGFfit} shows the parameters obtained.
It seems that $\chi^{2}/\mathrm{DoF}$ shown in table~\ref{table:gTGFfit} are rather large, especially for $L/a=12$.
This is caused by using wider range than needed for the fitting.
What we need is a smooth interpolating formula in the scaling region but not the fitting itself so 
we do not have to take the value of the $\chi^{2}/\mathrm{DoF}$ seriously.
The scaling region of figure~\ref{figure:gTGFfit} is magnified in figure~\ref{figure:gTGFfit_en} 
showing a smooth interpolation of the fitting.
Solving $g^{2}_{\TGF}(1/\Lmax,\beta^{\ast}) = u^{\ast}$ at each $u^{\ast}$ for $\beta^{\ast}$ 
using eq.~(\ref{eq:interplating_gTGF}), we obtain $\beta^{\ast}$ as shown 
in table~\ref{table:renoBareCoupling} in appendix~\ref{app:C}.

\begin{table}[t]
	\centering
	\begin{tabular}{|ccccccc|}
		\hline
		$L/a$ & $c_{1}$ & $c_{2}$ & $c_{3}$ & $c_{4}$ & $c_{5}$ & $\chi^{2}/\mathrm{DoF}$ 
		\\ \hline
		12 & $-$5.79(18) & 26.97(90) & $-$54.0(1.7) & 47.7(1.4) & $-$15.76(44) & 33.8(3.7)
		\\
		16 & $-$6.30(67) & 30.1(3.4) & $-$61.3(6.5) & 55.1(5.5) & $-$18.5(1.7) & 2.1(1.0)
		\\
		18 & $-$2.42(89) & 9.7(4.6) & $-$21.9(8.8) & 21.6(7.5) & $-$8.0(2.4) & 6.4(1.7)
		\\ \hline
	\end{tabular}
	\caption{Fitted parameters for eq.~(\ref{eq:interplating_gTGF}) at each lattice size.}
	\label{table:gTGFfit}
\end{table}

\begin{figure}[t]
	\centering
	\includegraphics[scale=\figscale,clip,angle=-90]{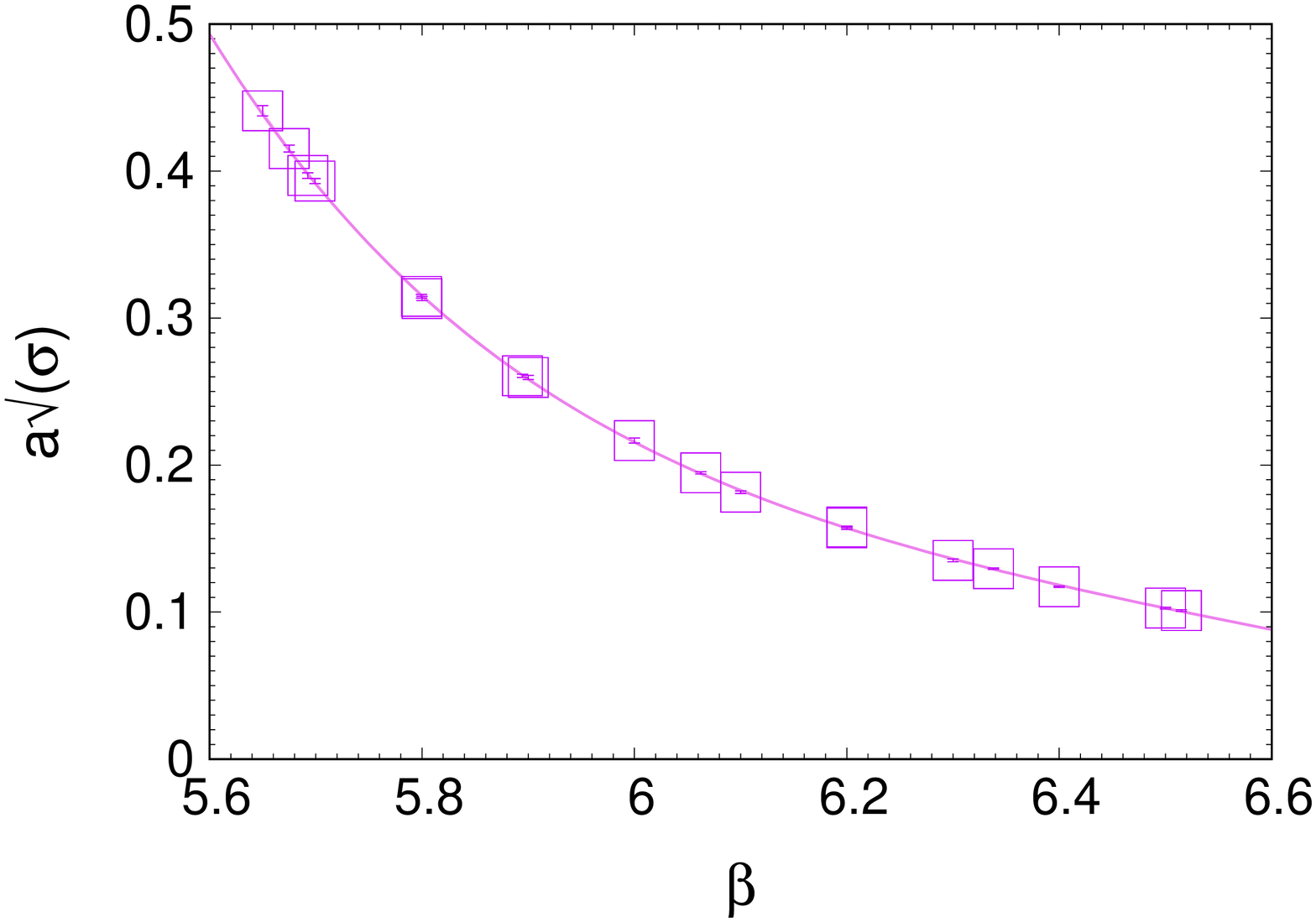}
	\caption{The $\beta$ dependence of the string tension $a\sqrt{\sigma}$.}
	\label{fig:interpolate_sigma}
\end{figure}

\begin{figure}[tb]
  \begin{tabular}{cc}
    \begin{minipage}{0.5\hsize}
      \centering
      \includegraphics[scale=0.35,clip,angle=-90]{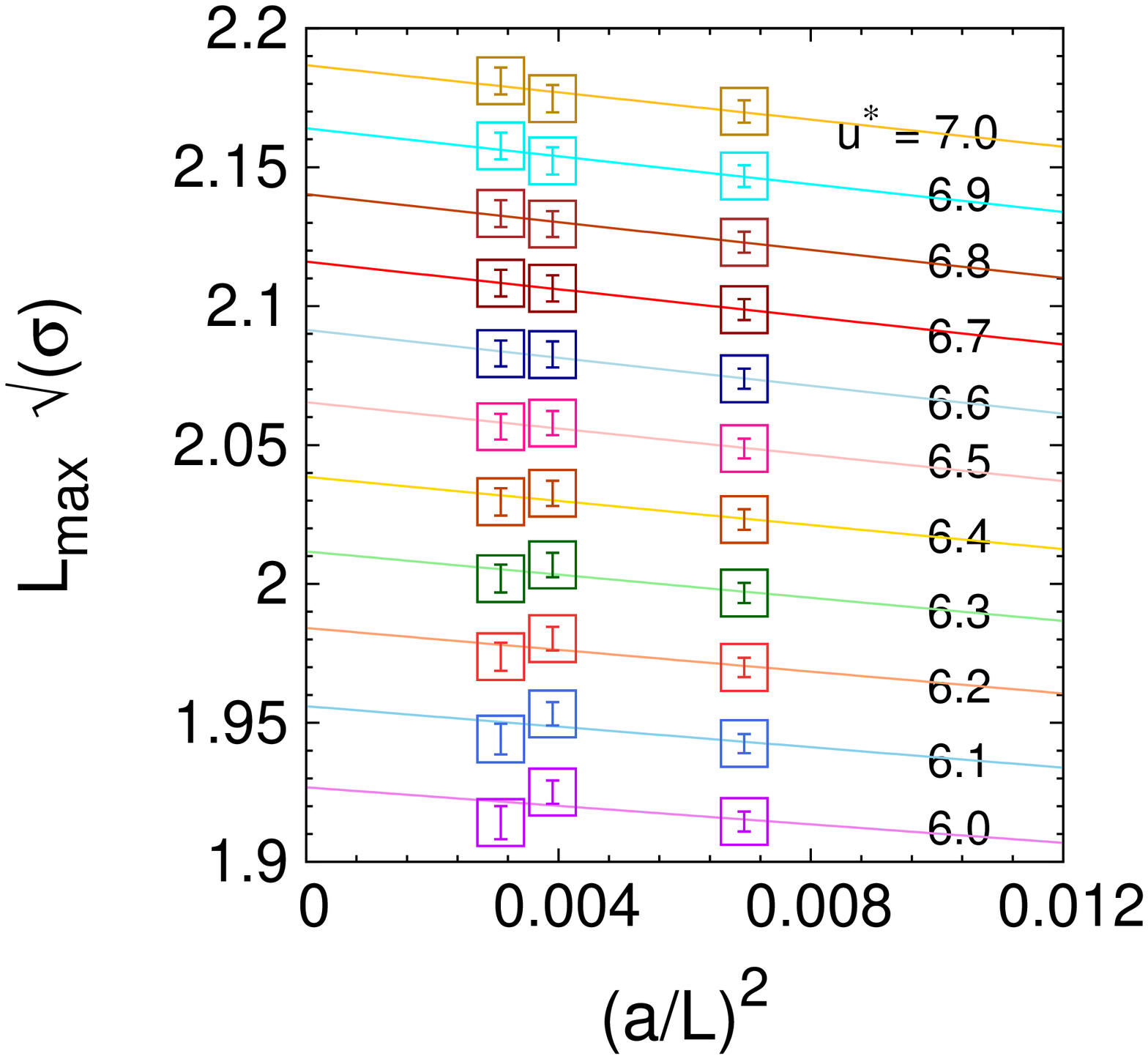}
    \end{minipage}
    \begin{minipage}{0.5\hsize}
      \centering
      \includegraphics[scale=0.35,clip,angle=-90]{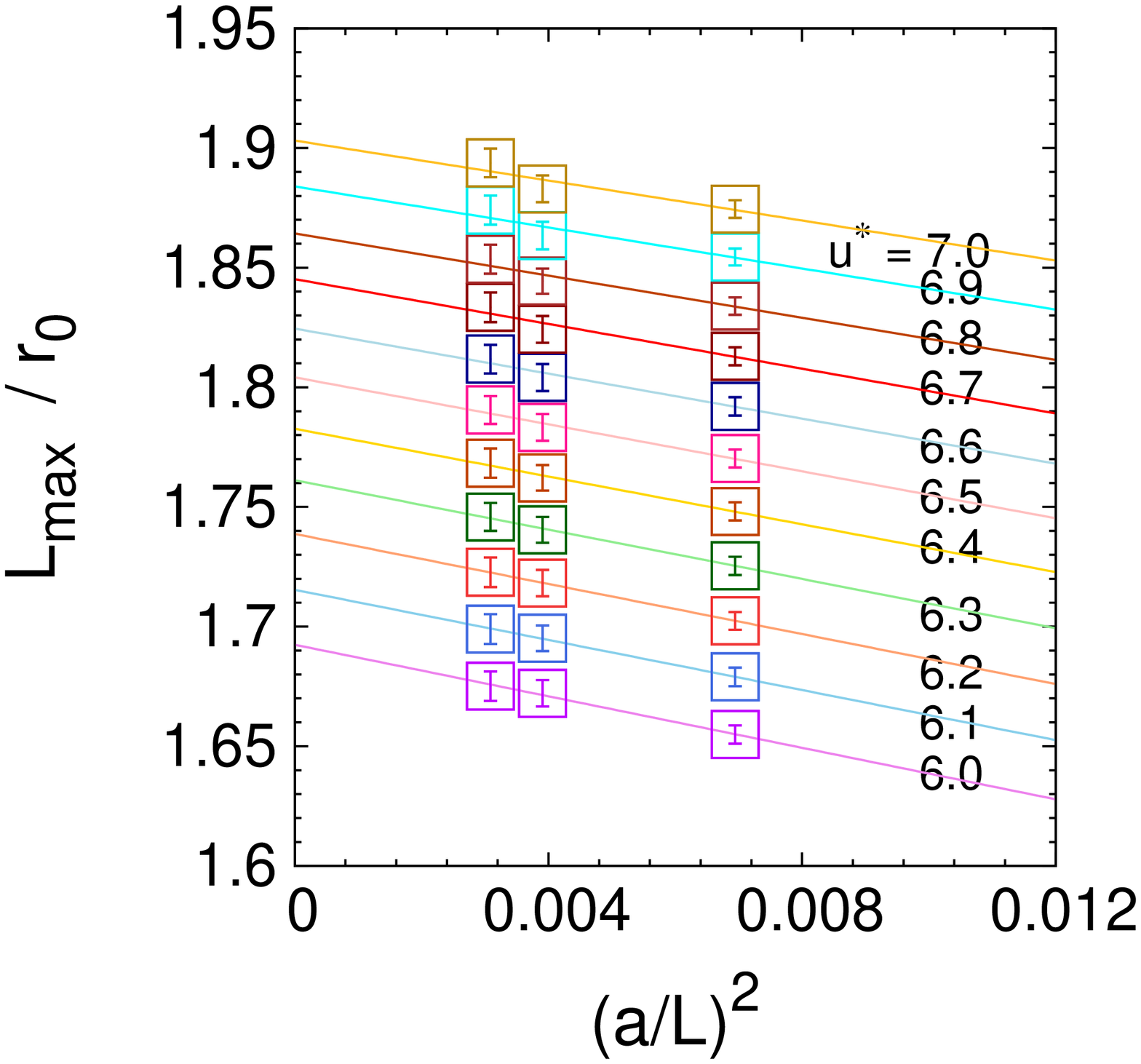}
    \end{minipage}
  \end{tabular}
  \caption{$(a/L)^2$ dependence of $\Lmax \sqrt{\sigma}$ (left) and $\Lmax /r_{0}$ (right) for each $u^{\ast}$. 
  The lines show the liner extrapolation to the continuum limit.}
  \label{figure:contST}
\end{figure}

Interpolating the data from refs.~\cite{Allton/Teper/Trivini:StringTension,Antonio/Okawa:StringTension} 
for $a\sqrt{\sigma}$ as a function of $\beta$, 
we obtain 
\begin{align}
  &(a\sqrt{\sigma})(\beta)=f(g_{0}^{2})\left(21935(1683)-10256(829)\beta+1608(136)\beta^{2}-84.2(7.4)\beta^{3}\right),
\notag \\
  &f(x)=\left(b_{0}x\right)^{-\frac{b_{1}}{2b_{0}^{2}}}\exp\left[-\frac{1}{2b_{0}x}\right],
  \label{int:ST}
\end{align}
with $\chi^{2}/{\mathrm{DoF}}\simeq1.43$. 
As plotted in figure~\ref{fig:interpolate_sigma}, eq.~(\ref{int:ST}) smoothly interpolates 
the data in the scaling region $\beta\in[6.11,6.515]$.
  Substituting $\beta^{\ast}$ from table~\ref{table:renoBareCoupling} into eq.~(\ref{int:ST}), 
and multiplying $\Lmax/a^{\ast}$ which corresponds to $\beta^{\ast}$ on it, 
we obtain  $\Lmax\sqrt{\sigma}$ at each $u^{\ast}$.
Table~\ref{table:renoST} in appendix~\ref{app:C} shows the values of $\Lmax \sqrt{\sigma}$ 
before taking the continuum limit.
The cut-off dependence of $\Lmax\sqrt{\sigma}$ for each $u^{\ast}$ is shown in the left panel of 
figure~\ref{figure:contST}. 
The values in the continuum limit are tabulated in the middle column of table~\ref{table:STSS}.

\begin{figure}[t]
	\centering
	\includegraphics[scale=\figscale,clip,angle=-90]{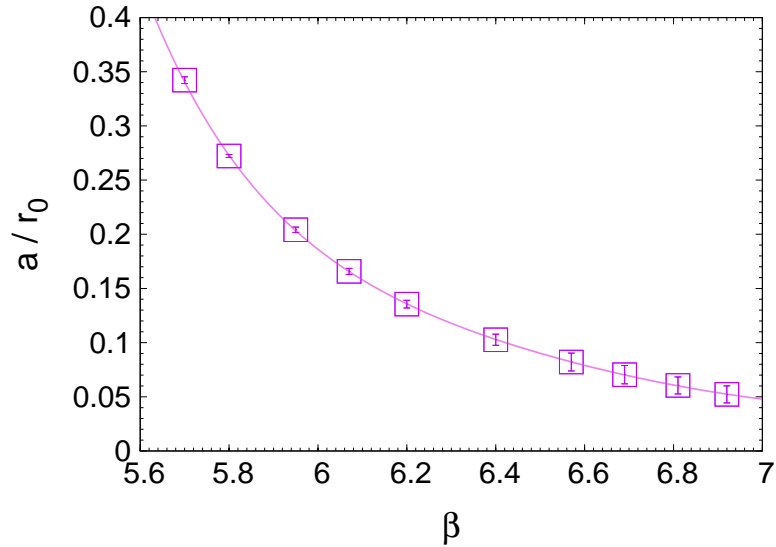}
	\caption{The $\beta$ dependence of the Sommer scale $a/r_{0}$.}
	\label{fig:interpolate_r0}
\end{figure}

\begin{table}[t]
	\centering
	\begin{tabular}{|ccc|}
		\hline	
		$u^{\ast}=g_{\TGF}^{2}(1/\Lmax )$ & $\Lmax \sqrt{\sigma}$ & $\Lmax /r_{0}$
		\\ \hline
		6.0                                         & 1.9268(81)                      & 1.6924(92)
		\\
		6.1                                         & 1.9560(78)                      & 1.7154(92)
		\\
		6.2                                         & 1.9842(76)                      & 1.7387(92)
		\\
		6.3                                         & 2.0117(76)                      & 1.7611(89)
		\\
		6.4                                         & 2.0386(77)                      & 1.7827(90)
		\\
		6.5                                         & 2.0654(73)                      & 1.8042(89)
		\\
		6.6                                         & 2.0914(75)                      & 1.8245(90)
		\\
		6.7                                         & 2.1160(76)                      & 1.8452(92)
		\\
		6.8                                         & 2.1403(77)                      & 1.8643(89)
		\\
		6.9                                         & 2.1640(78)                      & 1.8840(91)
		\\
		7.0                                         & 2.1867(79)                      & 1.9031(90)
		\\ \hline
	\end{tabular}
	\caption{$\Lmax \sqrt{\sigma}$ and  $\Lmax /r_{0}$ for each $u^{\ast}$ in the continuum limit.}
	\label{table:STSS}
\end{table}

We analyze $\Lmax /r_{0}$ similarly to the case of $\Lmax \sqrt{\sigma}$.
The interpolating formula is 
\begin{align}
 \frac{a}{r_{0}}(\beta) = & f(g_{0}^{2})
 \notag\\
  &\times\left(71325(12239)-43166(7824)\beta+9815(1873)\beta^{2}-992(199)\beta^{3}+37.6(7.9)\beta^{4}\right)
\end{align}
with $\chi^{2}/\mathrm{DoF}\simeq1.76$ (figure~\ref{fig:interpolate_r0} shows the interpolation in the scaling region).
We list the values of $\Lmax /r_{0}$ at each renormalization condition $u^{\ast}$ 
with finite lattice cut-off in table~\ref{table:renoSS} in appendix~\ref{app:C}.
The cut-off dependence and the values in the continuum limit are shown in the right panel 
of figure~\ref{figure:contST} and the right column in table~\ref{table:STSS}, respectively.


\section{$\Lambda$-parameter ratio $\LambdaSF/\LambdaTGF$ and $\Lambda_{\MSbar}$}
\label{sec:ratio}

To move from the TGF scheme to the $\MSbar$ scheme, we need the $\Lambda$-parameter ratio $\LambdaMSbar/\LambdaTGF$. 
Usually the ratio is calculated with the one-loop perturbation theory but 
the value is not yet available at the present time, while there is an ongoing project~\cite{bribian} of the perturbative calculation.
As we already know the ratio $\LambdaMSbar/\LambdaSF$~\cite{Sint/Sommer:LambdaSF}, 
what we have to calculate is the ratio $\LambdaSF/\LambdaTGF$. 
Since both $g_{\SF}^{2}$ and $g_{\TGF}^{2}$ can be evaluated on the lattice with the same cut-off and 
with the renormalization scale (that is, $a$ and $L$ are the same), 
we can evaluate them with the Monte Carlo simulation on the lattice.
We employ the two-loop formula~\cite{SFREFS,SF:two-loop},
\begin{equation}
  c_{\mathrm{t}}(g_{0})=1-0.08900(5)g_{0}^{2}-0.0294(3)g_{0}^{4},
\end{equation}
for the $O(a)$-improvement boundary correction in the SF simulations 
so that $g_{\SF}^{2}$ is $O(a)$-improved at the two-loop level.

Let us denote the SF and TGF couplings at the gauge coupling $\beta$ 
on a finite box $(L/a)^{4}$ by $g_{\SF}^{2}(a/L,\beta)$ and $g_{\TGF}^{2}(a/L,\beta)$, respectively. 
In a weak coupling region, these couplings are related through 
\begin{align}
  \frac{g_{\SF}^{2}(a/L,\beta)}{g_{\TGF}^{2}(a/L,\beta)} = 1 + c_{\mathrm{g}}(a/L)g_{\TGF}^{2}(a/L,\beta)+\cdots .
  \label{ratio:couplings}
\end{align}
We extract the value of $c_g(a/L)$ by investigating 
$g_{\TGF}^{2}(a/L,\beta)$ dependence of the ratio~(\ref{ratio:couplings}).
Both couplings $g_{\TGF}^{2}$ and $g_{\SF}^{2}$ are numerically evaluated at $\beta=40$, $60$ 
and $80$ on $L/a=8$, $10$, $12$ and $16$ lattices.
Since the TGF scheme is automatically free from $O(a)$ errors and $g_{\SF}$ is $O(a)$-improved, 
the $a/L$ dependence of $c_{\mathrm{g}}(a/L)$ should be
\begin{align}
  c_{\mathrm{g}}(a/L) = c_{\mathrm{g}}^{(0)}+c_{\mathrm{g}}^{(1)}\left(\frac{a}{L}\right)^{2}+\cdots .
  \label{coef'c_g}
\end{align}
The ratio of the $\Lambda$-parameters is defined by
\begin{equation}
	\frac{\LambdaSF}{\LambdaTGF}=c\times\exp\left[\frac{c_{\mathrm{g}}^{(0)}}{2b_{0}}\right].
	\label{LambdaRatio}
\end{equation}
with $c_{\mathrm{g}}^{(0)}$ from the continuum limit of $c_{\mathrm{g}}(a/L)$.

\begin{figure}[H]
  \begin{tabular}{c}
    \begin{minipage}{1.0\hsize}
      \centering
      \includegraphics[scale=\figscale,clip,angle=-90]{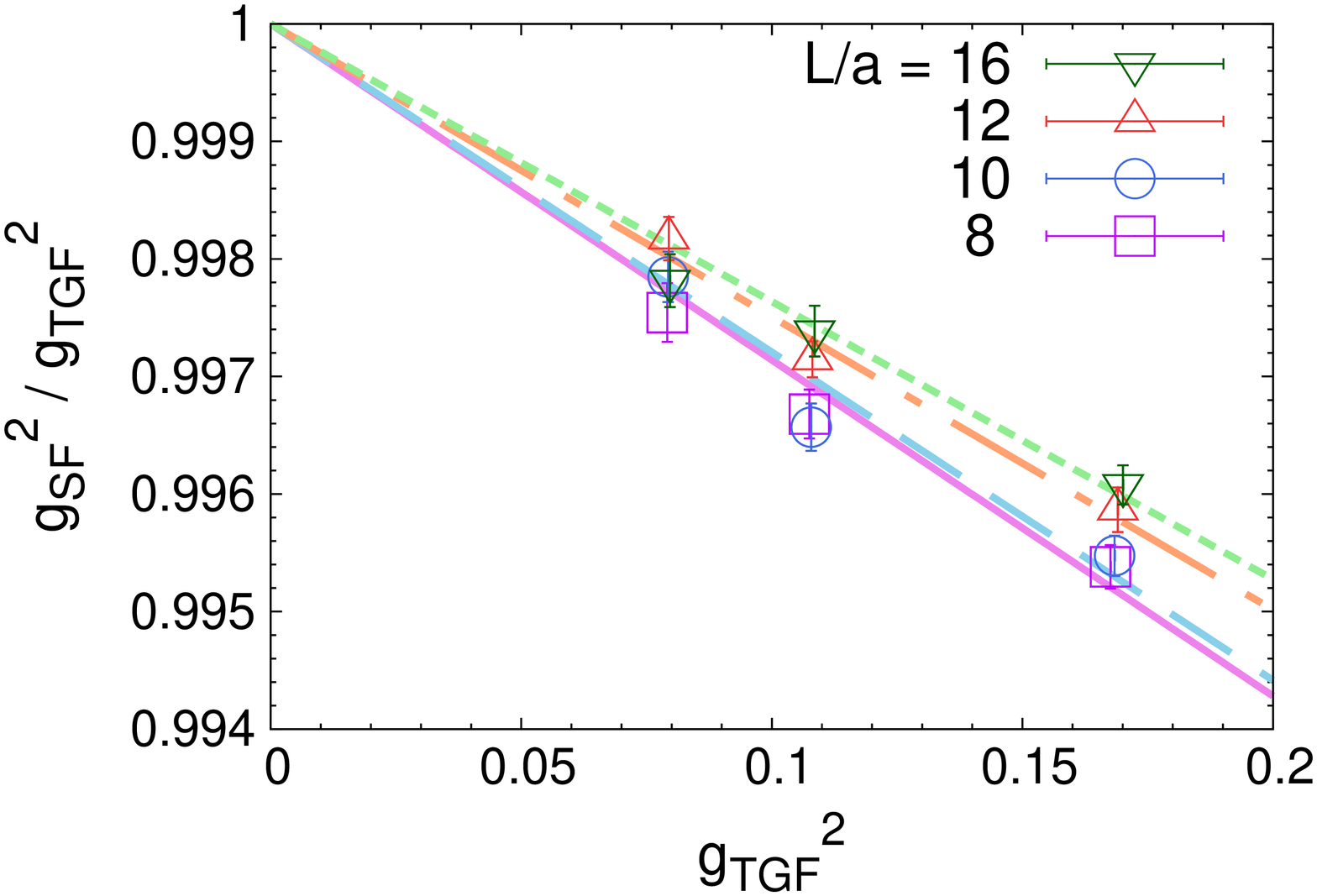}
      \caption{The ratio between the SF coupling and TGF coupling vs the TGF coupling. 
               The lines show the fit results with linear fitting.}
      \label{figure:TGFandSFcoupilngratio}
    \end{minipage}
  \\ \ \\
    \begin{minipage}{1.0\hsize}
      \centering
      \includegraphics[scale=\figscale,clip,angle=-90]{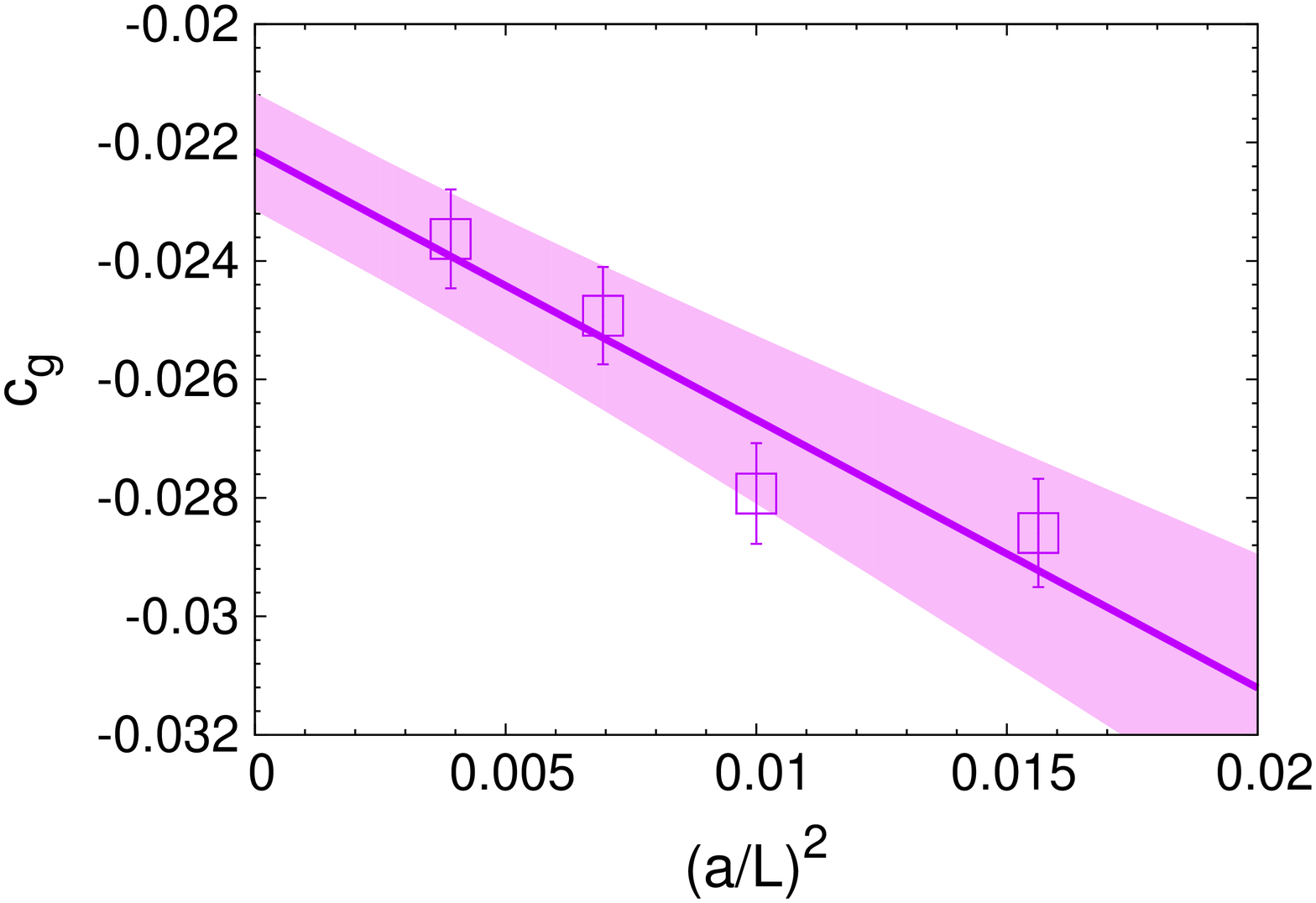}
      \caption{The coefficient $c_{\mathrm{g}}(a/L)$ vs $(a/L)^{2}$.}
      \label{figure:c}
    \end{minipage}
  \end{tabular}
\end{figure}

In table~\ref{table:data;weakcoupling} we list the TGF and SF couplings measured on each lattice size and each $\beta$. 
Figure~\ref{figure:TGFandSFcoupilngratio} shows $g_{\SF}^{2}(a/L,\beta)/g_{\TGF}^{2}(a/L,\beta)$ 
as a function of $g_{\TGF}^{2}(a/L,\beta)$. 
We fit the data linearly in $g_{\TGF}^{2}$ and the lines drawn in the figure are the fit results. 
Table~\ref{table:couplingratio-results} summarizes the fitted value of $c_{\mathrm{g}}(a/L)$ for each $L/a$. 
In figure~\ref{figure:c}, we plot $c_{\mathrm{g}}(a/L)$ as a function of $(a/L)^{2}$. 
Fitting the data linearly in $(a/L)^{2}$,
we obtain
\begin{equation}
	c_{\mathrm{g}}^{(0)}=-0.02215(99)
	\label{result:c_g}
\end{equation}
with $\chi^{2}/\mathrm{DoF}\simeq1.48$.
Consequently, eq.~(\ref{LambdaRatio}) with $c=0.3$ yields
\begin{equation}
	\frac{\LambdaSF}{c\LambdaTGF}=0.8530(61),
	\label{result:LambdaRatio}
\end{equation}
where the error quoted is the statistical one.

\begin{table}[tb]
	\centering
	\begin{tabular}{|cccc|cccc|}
		\hline
		$L/a$ & $\beta$ & $g_{\TGF}^{2}$ & $g_{\SF}^{2}$ & $L/a$ & $\beta$ & $g_{\TGF}^{2}$ & $g_{\SF}^{2}$
		\\ \hline
		 8    & 40      & 0.167587(25)       & 0.166813(18)      & 12    & 40      & 0.169048(26)       & 0.168350(19)
		\\
		      & 60      & 0.107511(18)       & 0.107154(13)      &       & 60      & 0.108094(14)       & 0.1077858(82)
		\\
		      & 80      & 0.079132(18)       & 0.0789374(73)     &       & 80      & 0.079439(11)       & 0.079294(10)
		\\ \hline
		10    & 40      & 0.168404(22)       & 0.167642(19)      & 16    & 40      & 0.170093(21)       & 0.169426(19)
		\\
		      & 60      & 0.107848(16)       & 0.107478(15)      &       & 60      & 0.108526(19)       & 0.108242(13)
		\\
		      & 80      & 0.079311(15)       & 0.0791399(81)     &       & 80      & 0.079700(16)       & 0.0795263(79)
		\\ \hline
	\end{tabular}
	\caption{TGF and SF couplings on each lattice and each $\beta$ in the weak coupling region.}
	\label{table:data;weakcoupling}
\end{table}

\begin{table}[H]
	\centering
	\begin{tabular}{|ccc|ccc|}\hline
		$L/a$ & $c_{\mathrm{g}}(L/a)$ & $\chi^{2}/\mathrm{DoF}$ & $L/a$      & $c_{\mathrm{g}}(L/a)$ & $\chi^{2}/\mathrm{DoF}$
		\\ \hline
		 8    & $-$0.02859(92)        & 1.42                    & 12         & $-$0.02492(82)        & 0.98
		\\
		10    & $-$0.02793(85)        & 2.76                    & 16         & $-$0.02363(84)        & 1.11
		\\ 
		\hline
	\end{tabular}
	\caption{The fit results for $c_{\mathrm{g}}$ at each lattice.}
	\label{table:couplingratio-results}
\end{table}

\begin{figure}[t]
      \centering
      \includegraphics[scale=\figscale,clip,angle=-90]{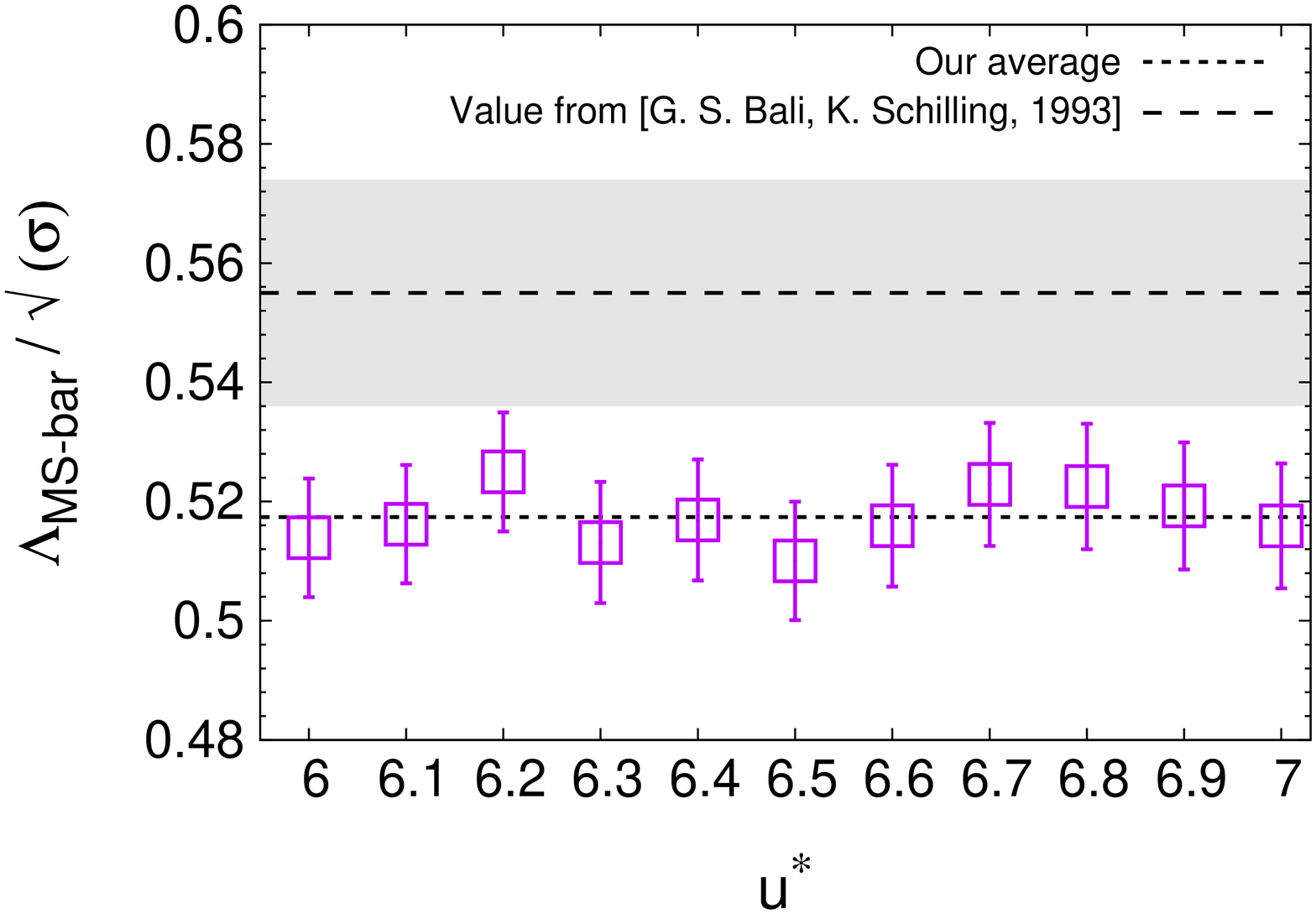}
      \caption{The intermediate scale ($u^{\ast}=g_{\TGF}^{2}(1/\Lmax)$) dependence of our results $\LambdaMSbar/\sqrt{\sigma}$. 
               The dotted lines are average over our results. 
               The dashed lines are the known values
               $\LambdaMSbar/\sqrt{\sigma}=0.555(^{+19}_{-17})$~\cite{Bali/Schilling:LambdaMSbar-StringTension}
               and the gray band denotes $1\sigma$.}
      \label{fig:final1}
\end{figure}

\begin{figure}[t]
      \centering
      \includegraphics[scale=\figscale,clip,angle=-90]{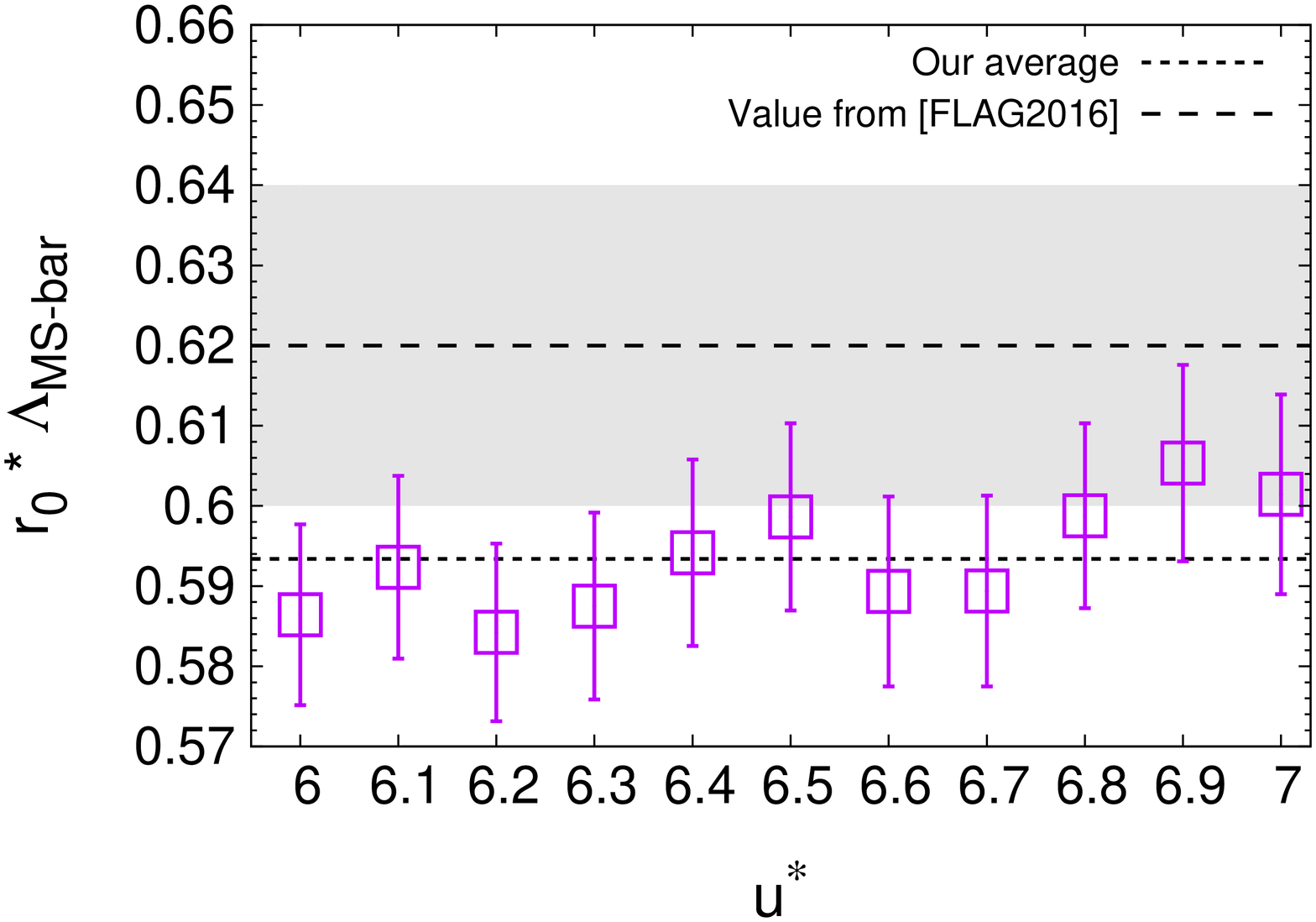}
      \caption{Same as figure~\ref{fig:final1}, but for $r_{0}\LambdaMSbar$.
               The dashed lines are the known values $r_{0}\LambdaMSbar=0.62(2)$~\cite{FLAG}.}
      \label{fig:final}
\end{figure}

We can now evaluate $\Lambda_{\MSbar}$ according to our strategy eq.~(\ref{strategy}).
We assemble $\LambdaSF/\LambdaMSbar=0.48811(1)$~\cite{Sint/Sommer:LambdaSF} 
and the results for $\Lmax \LambdaTGF$, $\Lmax A_{\mathrm{phys}}$, and $\LambdaSF/\LambdaTGF$ 
(tables~\ref{table:LambdaTGF},~\ref{table:STSS} and eq.~(\ref{result:LambdaRatio}), respectively).
Figures~\ref{fig:final1} and~\ref{fig:final} show the renormalization condition $u^{\ast}$ dependence of 
$\LambdaMSbar/\sqrt{\sigma}$ and $r_{0}\LambdaMSbar$, respectively.
In these figures, the square symbols with error bar, which is statistical one, are our results and 
the dotted line is the average over our results with different $u^{\ast}$.  
The dashed lines with gray band are 
from refs.~\cite{Bali/Schilling:LambdaMSbar-StringTension} and~\cite{FLAG} for comparison.
We observe no renormalization condition dependence as expected. 
Our final estimates are
\begin{align}
  \frac{\LambdaMSbar}{\sqrt{\sigma}} &= 0.517(10)_{\mathrm{stat.}}(^{+8}_{-7})_{\mathrm{syst.}},
 \label{result:st}
\\
                   r_{0}\LambdaMSbar &= 0.593(12)_{\mathrm{stat.}}(^{+12}_{-9})_{\mathrm{syst.}}.
 \label{result:ss}
\end{align}
The central values are from the averages stated above.
The systematic error is estimated from the renormalization condition dependence.
Our results of $\LambdaMSbar/\sqrt{\sigma}$ and $r_{0}\LambdaMSbar$ are 
compatible within $1.8\sigma$ and $1.1\sigma$ with 
the known values
     $\LambdaMSbar/\sqrt{\sigma} = 0.555(^{+19}_{-17})$
   from~\cite{Bali/Schilling:LambdaMSbar-StringTension}
          and $r_{0}\LambdaMSbar = 0.62(2)$
   from~\cite{FLAG}, respectively.



\section{Summary}
\label{sec:summary}

We have evaluated the $\Lambda$-parameter in the $\overline{\mathrm{MS}}$ scheme for 
the pure SU(3) gauge theory via the twisted gradient flow method according to our strategy shown in~(\ref{strategy}). 
Our results are summarized in eqs.~(\ref{result:st}) and~(\ref{result:ss}). 
To obtain the results we have determined the $\Lambda$-parameter ratio between the TGF scheme 
and the SF scheme with lattice simulations, which is a non-trivial step in our analysis.
Having obtained sufficiently close values to the known ones in eqs.~(\ref{result:st}) and~(\ref{result:ss}), 
we verified the ratio $\Lambda_{\SF}/\Lambda_{\TGF}$ (\ref{result:LambdaRatio}) 
determined with non-perturbative simulations. 
To further confirm the value of the ratio $\Lambda_{\SF}/\Lambda_{\TGF}$, 
it would be interesting to compare our ratio with the analytic one from the explicit 
perturbative calculation~\cite{bribian}.


\acknowledgments

The numerical simulations have been done on the INSAM (Institute for Nonlinear Sciences and Applied Mathematics) 
cluster system at Hiroshima University. 
This work was partly supported by JSPS KAKENHI Grant Numbers 26400249 and 16K05326.
I. K. is supported by MEXT as ``Priority Issue on Post-K computer'' (Elucidation of the Fundamental Laws and Evolution of the Universe) and JICFuS, by which K.-I. I. is also partially supported.


\newcommand{\tr}{\mathop{\mathrm{tr}}}

\appendix


\section{Derivation of the action}
\label{app:A}

In this appendix, we derive the action with periodic variables~(\ref{WilsonAction}). 
We start from the following action in SU($N_{\mathrm{C}}$) defined on a $(L/a)^{4}\equiv\hat{L}^{4}$ 
lattice with the twisted boundary condition on the $x$-$y$ plane and periodic boundary condition in $z$ and $t$ directions:
\begin{align}
  S = \frac{\beta}{2N_{\mathrm{C}}}\sum_{{n,\mu,\nu}\atop{(\mu\neq\nu)}}\mathrm{Tr}\left[P_{\mu\nu}[n;V]\right],
\end{align}
where
\begin{align}
	P_{\mu\nu}[n;V] = V_{\mu}(n)V_{\nu}(n+\hat{\mu}) V_{\mu}^{\dagger}(n+\hat{\nu}) V_{\nu}^{\dagger}(n)
\end{align}
is a plaquette variable made of link variables $V_{\mu}(n)$ with the twisted boundary condition:
\begin{align}
	V_{\mu}(n+\hat{L}\hat\nu)&=\Gamma_{\nu}V_{\mu}(n)\Gamma_{\nu}^{\dagger} &(\nu=1,2),
	\\
	V_{\mu}(n+\hat{L}\hat\nu)&=V_{\mu}(n)                                   &(\nu=3,4),
\end{align}
where $N_{\mathrm{C}}\times N_{\mathrm{C}}$ unitary matrix $\Gamma_{\nu}$ ($\nu=1,2$) is called twist matrix and satisfies
\begin{align}
    \Gamma_{1}\Gamma_{2}&=\omega\Gamma_{2}\Gamma_{1}, 
  \qquad
    \omega =\exp\left[\frac{2\pi i}{N_{\mathrm{C}}}\right].
\end{align}

Let us eliminate the link variables on $n_{1}=0$ or $n_{2}=0$ by using the variables on $n_{i}=\hat{L}$. 
The plaquette on $n_{1}=n_{2}=0$ becomes
\begin{align}
	\mathrm{Tr}P_{12}[(0,0,n_{3},n_{4});V]=\omega^{\ast}\mathrm{Tr}[&V_{1}(\hat{L},\hat{L},n_{3},n_{4})\Gamma_{1}V_{2}(1,\hat{L},n_{3},n_{4})\Gamma_{2}
	\nonumber\\
	&\Gamma_{1}^{\dagger}V_{1}^{\dagger}(\hat{L},1,n_{3},n_{4})\Gamma_{2}^{\dagger}V_{2}^{\dagger}(\hat{L},\hat{L},n_{3},n_{4})]. 
\end{align}
By introducing the following new variables for $n_{1,2}=1,2,\dots,\hat{L}$
\begin{align}
	U_{1}(\hat{L},n_{2},n_{3},n_{4})&\equiv V_{1}(\hat{L},n_{2},n_{3},n_{4})\Gamma_{1}=\Gamma_{1}V_{1}(0,n_{2},n_{3},n_{4}),
	\\
	U_{2}(n_{1},\hat{L},n_{3},n_{4})&\equiv V_{2}(n_{1},\hat{L},n_{3},n_{4})\Gamma_{2}=\Gamma_{2}V_{2}(n_{1},0,n_{3},n_{4}),
	\\
	U_{\mu}(n_{1},n_{2},n_{3},n_{4}) &\equiv V_{\mu}(n_{1},n_{2},n_{3},n_{4})\quad\text{for others},
\end{align}
it becomes
\begin{align}
	\mathrm{Tr}P_{12}[(0,0,n_{3},n_{4});V]=\omega^{\ast}\mathrm{Tr}[&U_{1}(\hat{L},\hat{L},n_{3},n_{4})U_{2}(1,\hat{L},n_{3},n_{4})
	\nonumber \\
	&U_{1}^{\dagger}(\hat{L},1,n_{3},n_{4})U_{2}^{\dagger}(\hat{L},\hat{L},n_{3},n_{4})].
\end{align}
Except for the overall factor $\omega$, this is exactly the plaquette with \emph{periodic} link variables $U_{\mu}(n)$. 
Therefore we define link variables on $n_{1}=0$ and $n_{2}=0$ through the periodic boundary condition:
\begin{align}
	U_{\mu}(0,n_{2},n_{3},n_{4})&\equiv U_{\mu}(\hat{L},n_{2},n_{3},n_{4}),
	\\
	U_{\mu}(n_{1},0,n_{3},n_{4})&\equiv U_{\mu}(n_{1},\hat{L},n_{3},n_{4}),	
	\\
	U_{\mu}(0,0,n_{3},n_{4})&\equiv U_{\mu}(\hat{L},\hat{L},n_{3},n_{4}).
\end{align}
Similar calculations show other plaquettes become those with $U_{\mu}(n)$ without overall factor. 
Then, we finally obtain the action with periodic link variable $U_{\mu}(n)$
\begin{align}
	S=\frac{\beta}{2N_{\mathrm{C}}}\sum_{{n,\mu,\nu}\atop{(\mu\neq\nu)}}\mathrm{Tr}\left[Z_{\mu\nu}(n)P_{\mu\nu}[n;U]\right],
\end{align}
where $Z_{\mu\nu}(n)=Z_{\nu\mu}^{\ast}(n)$ is given as
\begin{align}
	Z_{\mu\nu}(n)=
	\begin{cases}
		\omega^{\ast} & \mu=1,\nu=2,\ \text{and}\ n_{1}=n_{2}=0,
		\\
		1              & \text{otherwise}.
	\end{cases}
\end{align}


\section{The number of the configurations for $g_{\TGF}^{2}$}
\label{app:B}

We list the number of the configurations used to calculate $g_{\TGF}^{2}$ in table~\ref{table:number_of_configs_with_ac}.

\begin{table}[H]
	\centering
	\begin{tabular}{|r|rrrrrr|}
		\hline 
		& \multicolumn{6}{|c|}{number of configurations [autocorrelation length]}
		\\
		\multicolumn{1}{|c|}{$\beta $} & \multicolumn{1}{c}{12}   & \multicolumn{1}{c}{16}   & \multicolumn{1}{c}{18}   & \multicolumn{1}{c}{24}   & \multicolumn{1}{c}{36}   & \multicolumn{1}{@{}l|}{$L/a$}
		\\ \hline
		 6.11   & 91300[1.7] &            &            &            &            & 
		\\
		 6.20   & 69500[2.3] &            &            &            &            & 
		\\
		 6.29   &            &            & 2750[2.5]  &            &            & 
		\\
		 6.30   &            & 27500[3.7] &            &            &            & 
		\\
		 6.38   &            &            & 16760[5.3] &            &            & 
		\\
		 6.40   & 19500[1.0] & 26000[8.8] & 15266[4.9] &            &            & 
		\\
		 6.50   & 15500[0.8] & 14300[8.7] & 14804[7.8] & 8746[4.8]  &            & 
		\\
		 6.60   &            & 5750[1.1]  &            & 9340[14.1] &            & 
		\\
		 6.75   & 15500[0.7] & 2500[1.0]  & 2200[1.2]  & 6000[2.8]  &            & 
		\\
		 6.90   &            &            &            & 5212[2.0]  & 4670[17.6] & 
		\\
		 7.00   & 15500[0.6] & 1600[1.3]  & 1420[0.9]  & 2556[1.6]  & 2552[12.0] & 
		\\
		 7.25   & 15500[0.6] & 1750[0.6]  & 1900[1.0]  & 2200[1.7]  & 800[2.4]   & 
		\\
		 7.50   & 15500[0.7] & 1300[0.6]  & 1350[1.0]  & 1420[1.7]  & 800[1.5]   & 
		\\
		 7.75   & 15500[0.7] & 1800[0.7]  & 1200[0.9]  & 1100[1.7]  & 800[2.3]   & 
		\\
		 8.00   & 15500[0.7] & 1100[0.7]  & 1200[0.9]  & 1200[1.2]  & 800[2.9]   & 
		\\
		 8.25   & 15000[0.7] & 1700[0.9]  & 1480[0.9]  & 1200[1.1]  & 800[1.6]   & 
		\\
		 8.50   & 15000[0.6] & 1400[0.7]  & 1300[1.0]  & 1200[1.0]  & 800[1.5]   & 
		\\
		 9.00   & 15500[0.7] & 1300[0.8]  & 1100[1.5]  & 880[1.2]   & 800[2.2]   & 
		\\
		 9.50   & 15500[0.9] & 1800[1.1]  & 1200[1.2]  & 780[1.7]   & 800[1.9]   & 
		\\
		10.00   & 15500[0.7] & 1800[0.7]  & 1600[0.7]  & 780[1.2]   & 800[2.4]   & 
		\\	\hline
	\end{tabular}
	\caption{Number of configurations after thermalization used to calculate $g^{2}_{\TGF}$.}
	\label{table:number_of_configs_with_ac}
\end{table}


\clearpage
\section{Tables to evaluate $\Lmax /A_{\mathrm{phys}}$}
\label{app:C}

In tables~\ref{table:renoBareCoupling},~\ref{table:renoST} and~\ref{table:renoSS} we collect values needed to evaluate $\Lmax /A_{\mathrm{phys}}$ in section~\ref{sec:STSS}.

\begin{table}[H]
	\centering
	\begin{tabular}{|c|cccc|}
		\hline
		           &             & $\beta^{\ast}$ &          & 
		\\
		$u^{\ast}$ & 12          & 16          & 18          & $\Lmax /a^{\ast}$
		\\ \hline
		6.0        & 6.18950(20) & 6.38859(55) & 6.47578(82) & 
		\\
		6.1        & 6.17971(20) & 6.37826(54) & 6.46493(79) & 
		\\
		6.2        & 6.17032(20) & 6.36834(54) & 6.45449(76) & 
		\\
		6.3        & 6.16130(20) & 6.35881(54) & 6.44445(73) & 
		\\
		6.4        & 6.15263(20) & 6.34965(54) & 6.43477(71) & 
		\\
		6.5        & 6.14429(20) & 6.34084(55) & 6.42545(68) & 
		\\
		6.6        & 6.13627(20) & 6.33235(57) & 6.41646(67) & 
		\\
		6.7        & 6.12854(21) & 6.32418(58) & 6.40779(65) & 
		\\
		6.8        & 6.12109(21) & 6.31630(60) & 6.39942(64) & 
		\\
		6.9        & 6.11391(22) & 6.30870(62) & 6.39133(64) & 
		\\
		7.0        & 6.10697(23) & 6.30137(65) & 6.38351(63) & 
		\\ \hline
	\end{tabular}
	\caption{The bare coupling $\beta^{\ast}$ at the renormalization condition $u^{\ast}$.}
	\label{table:renoBareCoupling}
\end{table}

\begin{table}[H]
	\centering
	\begin{tabular}{|c|cccc|}
		\hline
		           &\multicolumn{3}{c}{$(\Lmax /a^{\ast})\cdot(a^{\ast}\sqrt{\sigma})$}& 
		\\
		$u^{\ast}$ & 12         & 16         & 18         & $\Lmax /a^{\ast}$
		\\ \hline
		6.0        & 1.9145(36) & 1.9251(42) & 1.9141(59) & 
		\\
		6.1        & 1.9426(34) & 1.9532(42) & 1.9441(56) & 
		\\
		6.2        & 1.9700(35) & 1.9803(43) & 1.9738(51) & 
		\\
		6.3        & 1.9967(36) & 2.0068(44) & 2.0019(50) & 
		\\
		6.4        & 2.0232(37) & 2.0326(46) & 2.0296(49) & 
		\\
		6.5        & 2.0487(36) & 2.0579(44) & 2.0566(46) & 
		\\
		6.6        & 2.0739(36) & 2.0826(47) & 2.0829(47) & 
		\\
		6.7        & 2.0987(38) & 2.1064(47) & 2.1083(47) & 
		\\
		6.8        & 2.1230(38) & 2.1295(47) & 2.1333(49) & 
		\\
		6.9        & 2.1468(40) & 2.1522(49) & 2.1577(48) & 
		\\
		7.0        & 2.1701(40) & 2.1747(49) & 2.1811(48) & 
		\\ \hline
	\end{tabular}
        \caption{$\Lmax \sqrt{\sigma}$ at each $u^{\ast}$.}
	\label{table:renoST}
\end{table}

\begin{table}[H]
	\centering
	\begin{tabular}{|c|cccc|}
		\hline
		           &\multicolumn{3}{c}{$(\Lmax /a^{\ast})\cdot(a^{\ast}/r_{0})$}& 
		\\
		$u^{\ast}$ & 12          & 16          & 18         & $\Lmax /a^{\ast}$
		\\ \hline
		6.0        & 1.6549(38)  & 1.6721(56)  & 1.6751(61) &
		\\
		6.1        & 1.6790(39)  & 1.6951(54)  & 1.6990(62) & 
		\\
		6.2        & 1.7024(37)  & 1.7182(55)  & 1.7227(62) & 
		\\
		6.3        & 1.7254(38)  & 1.7404(54)  & 1.7458(58) & 
		\\
		6.4        & 1.7481(38)  & 1.7622(53)  & 1.7683(61) & 
		\\
		6.5        & 1.7702(37)  & 1.7833(56)  & 1.7905(58) & 
		\\
		6.6        & 1.7921(38)  & 1.8041(56)  & 1.8118(59) & 
		\\
		6.7        & 1.8130(38)  & 1.8242(56)  & 1.8334(62) & 
		\\
		6.8        & 1.8340(36)  & 1.8444(53)  & 1.8536(61) & 
		\\
		6.9        & 1.8545(35)  & 1.8634(58)  & 1.8740(61) & 
		\\
		7.0        & 1.8745(37)  & 1.8829(56)  & 1.8938(60) & 
		\\ \hline
	\end{tabular}
	\caption{$\Lmax /r_{0}$ at each $u^{\ast}$.}
	\label{table:renoSS}
\end{table}


\end{document}